\documentclass[aoas]{imsart}

\RequirePackage{amsthm,amsmath,amsfonts,amssymb}
\RequirePackage[authoryear]{natbib}
\RequirePackage[colorlinks,citecolor=blue,urlcolor=blue]{hyperref}
\RequirePackage{graphicx}

\usepackage{bbold}
\usepackage{fancyvrb}

\startlocaldefs
\theoremstyle{plain}

\newtheorem{theorem}{Theorem}[section]

\theoremstyle{definition}

\endlocaldefs

\begin{document}

\begin{frontmatter}
\title{Raking mortality rates across cause, population group and geography with uncertainty quantification}
\runtitle{Raking mortality rates with uncertainty quantification}

\begin{aug}
\author[A]{\fnms{Ariane}~\snm{Ducellier}\ead[label=e1]{ducela@uw.edu}},
\author[A]{\fnms{Alexander}~\snm{Hsu}\ead[label=e2]{owlx@uw.edu}},
\author[A]{\fnms{Parkes}~\snm{Kendrick}\ead[label=e3]{parkesk@uw.edu}},
\author[A]{\fnms{Bill}~\snm{Gustavson}\ead[label=e4]{billg@uw.edu}},
\author[A]{\fnms{Laura}~\snm{Dwyer-Lindgren}\ead[label=e5]{ladwyer@uw.edu}},
\author[A]{\fnms{Christopher}~\snm{Murray}\ead[label=e6]{cjlm@uw.edu}},
\author[A]{\fnms{Peng}~\snm{Zheng}\ead[label=e7]{zhengp@uw.edu}}
\and
\author[A]{\fnms{Aleksandr}~\snm{Aravkin}\ead[label=e8]{saravkin@uw.edu}}
\address[A]{Institute for Health Metrics and Evaluation (IHME),  Seattle, WA\printead[presep={ ,\ }]{e1,e2,e3,e4,e5,e6,e7,e8}}
\end{aug}
  
\begin{abstract}
The Global Burden of Diseases, Injuries, and Risk Factors Study (GBD) is the single largest and most detailed scientific effort ever conducted to quantify levels and trends in health. This global health model to estimate mortality rates and other health metrics is run at different scales, leading to large data sets of results for a global region and its different sub-regions, or for a cause of death and different sub-causes for example. These models do not necessarily lead to consistent data tables where, for instance, the sum of the number of deaths for each of the sub-regions is equal to the number of deaths for the global region. Raking is widely used in survey inference and global health models to adjust the observations in contingency tables to given marginals, in the latter case reconciling estimates between models with different granularities. The results of global health models usually associate to the point estimates an uncertainty, such as standard deviations or confidence intervals. In this paper, we propose an uncertainty propagation approach that obtains, at the cost of a single solve, nearly the same uncertainty estimates as computationally intensive Monte Carlo techniques that pass thousands of observed and marginal samples through the entire raking process. We introduce a convex optimization approach that provides a unified framework to raking extensions such as uncertainty propagation, raking with differential weights, raking with different loss functions in order to ensure that bounds on estimates are respected, verifying the feasibility of the constraints, raking to margins either as hard constraints or as aggregate observations, and handling missing data. 
\end{abstract}


\end{frontmatter}

\section{Introduction}
\label{sec:introduction}

The Global Burden of Diseases, Injuries, and Risk Factors Study (GBD,~\cite{GBD}) aims to quantify levels and trends in health by creating a unique platform to compare the magnitude of diseases, injuries, and risk factors between age groups, sexes, countries, regions, and time, to compare health progress of countries and to understand the leading causes of health loss that could potentially be avoided, such as high blood pressure, smoking, and household air pollution.

Many models to estimate the health metrics covered by the GBD studies are run at different scales, leading to large data sets of results for a global region and its different sub-regions, or for a cause of death and different sub-causes, for example. These models do not necessarily lead to consistent data tables where, for instance, the sum of the number of deaths for each of the sub-regions is equal to the number of deaths for the global region. In order to ascertain the consistency of the results across geography, population groups, and causes of death or disease, it is thus necessary to adjust the output of the models to make sure that all the marginal totals match the corresponding totals for the population. Raking is widely used in survey inference and global health models to adjust the observations in contingency tables to given marginals, in the latter case reconciling estimates between models with different granularities. 

\begin{figure}[h!]
\begin{center}
\includegraphics[width=0.7\textwidth]{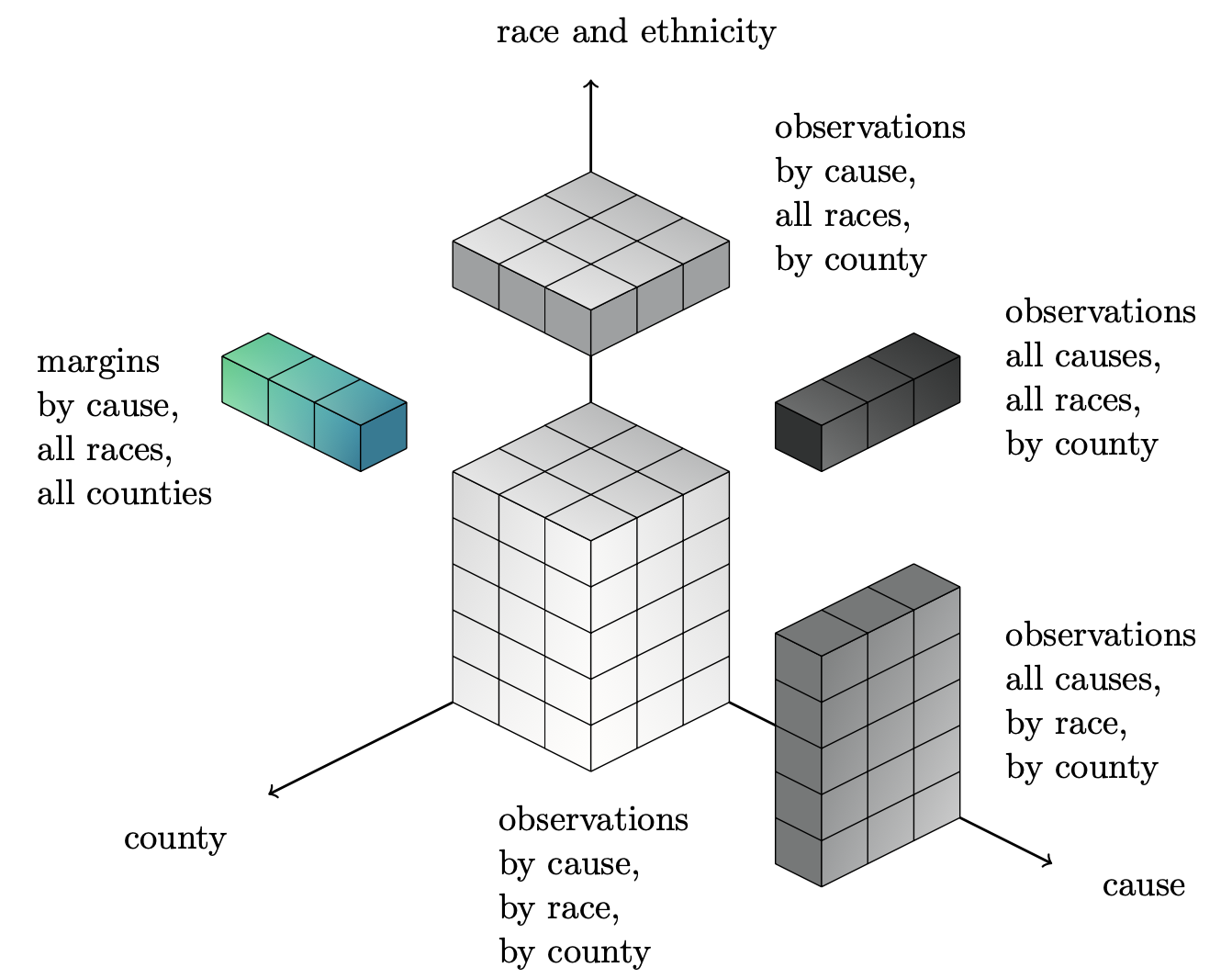}
\end{center}
\caption{Summary of the raking problem for the mortality estimates dataset.}
\label{fig:USHD_problem}
\end{figure}

In our specific application, age-standardized mortality by racial-ethnic group, county, and cause of death has been estimated using small area estimation models to death certificate data from the US National Vital Statistics system and population data from the US National Center for Health Statistics (\cite{DWY_2023}). The objective of the study is to understand how racial-ethnic disparities in the United States vary by location and cause of death. However, the outputs of the small-area estimation models are not consistent across cause of death, racial-ethnic group, and location. The data set contains the estimated number of deaths by cause, racial-ethnic group, and county for each US state, year, sex, and age group. These estimates need to be raked to aggregate estimates, typically all racial-ethnic groups state-wide cause-specific and all-cause counts, while trying to preserve key relational information in the detailed estimates, as described below and summarized in Figure~\ref{fig:USHD_problem}.

There are three different causes of death: communicable, maternal, neonatal, and nutritional diseases (Comm.), non-communicable diseases (NCD), and injuries (Inj.) and five different racial-ethnic groups: White, Black, American Indian or Alaska Native (AIAN), Asian or Pacific Islander (API), and Hispanic/Latino (Hisp.). Aggregate data are provided by the GBD (\cite{GBD_2021}) and are the total number of deaths for the state and the number of deaths due to each cause for the state. The margins must be used as constraints,  so the total number of deaths for the state is already equal to the sum of the number of deaths due to each cause for the state (shown as teal blocks in Figure~\ref{fig:USHD_problem}). On the other hand, all-causes aggregates by racial-ethnic group and county, all-causes all-racial-ethnic-groups aggregates by county, and all-racial-ethnic-groups cause-specific aggregates by county may not be consistent and can be used as observations to be raked rather than margins (shown as gray blocks in Figure~\ref{fig:USHD_problem}). 

In a previous study (\cite{DWY_2016}), the raking was not done on racial-ethnic-group. A follow-up study (\cite{DWY_2023}) used racial-ethnic-group but without adjusting the results of the models to the GBD margins. These studies needed only one- and two-dimensional raking to make all the outputs of the models consistent with each other. One-dimensional raking was done using a closed form solution (see Section 2 in Supplement 1) and the Iterative Proportional Fitting algorithm (IPF, ~\cite{DEM_1940} and ~\cite{STE_1942}) was used for the two-dimensional raking. Our main objective is to extend the specific solutions to the raking problems in~\cite{DWY_2016} and~\cite{DWY_2023} to a generalized raking problem for the next iteration of the GBD study, incorporating a range of extensions and functionality for improved accuracy and wider applicability.

Accounting for uncertainties in data input is crucial for the robustness and reliability of modern data analysis pipelines. By explicitly considering uncertainty, global health models provide more nuanced predictions with confidence intervals, aiding decision-makers in assessing risks and making more informed choices. The GBD studies used to take $1000$ samples from the posterior distribution of each health metric of interest, to get a reasonable representation of the mean, standard deviation and $2.5$ and $97.5$ percentiles of the distribution. This is also an efficient mechanism to capture covariance of health metrics across different categorical variables, such as age, sex, location, and year. However, as the number of health-metric outputs produced by the GBD studies models has increased, the data storage requirements and computational costs have also dramatically increased. Coupled with the generalized raking approach, we develop an uncertainty propagation approach that obtains, at the cost of a single solve, nearly the same uncertainty estimates as computationally intensive Monte Carlo techniques that pass thousands of observed and marginal samples through the entire raking process.

The main contribution of this paper is to formulate a generalized raking problem that encompasses multiple extensions, including propagating uncertainties through the raking problem via single solve, differential weighting to enable the uncertainty of inputs to affect the raking results, different loss functions (deviances) of practical interest (~\cite{DEV_1992}) including raking of bounded values, automated checks for the feasibility of the constraints, raking to margins either as hard constraints or as aggregate observations, and handling missing data. Several of these extensions, such as differential weighting and the inclusion of aggregate observations in the raking problem, were collected by~\cite{williams2024optimization}. A key contribution of the current paper is to bring to light the form and structure of the dual problem and to use it for improved understanding, fast algorithms, and more challenging cases. Duality is an essential ingredient in efficient algorithms, particularly for higher dimensions. The dual perspective also brings clarity on the impact of missing data and allows new methods of uncertainty propagation. 

We describe the generalized raking problem in Section~\ref{sec:methods} and use the implicit function theorem to design a specialized delta method for raking, efficiently propagating uncertainty from both observations and aggregates. We then describe how this generalized formulation provides practical functionalities for the user in Section~\ref{sec:discussion}. The application to the mortality rate estimates across cause, population group and geography is described in Section~\ref{sec:application}. Additional historical context on raking in the survey sampling domain are given in Supplement 1. Convex preliminaries and technical details and proofs about the raking methods are given in Supplement 2. Illustrations of the raking extensions on simple examples are given in Supplement 3. All of the methods are implemented in an open source Python package with an intuitive user interface, installable from PyPi\footnote{https://pypi.org/project/raking/}. We explain the interface to the open source tool in Supplement 4. 

\section{Methods}
\label{sec:methods}

\subsection{Raking under multiple constraints}
\label{sec:raking}

We first summarize the solution proposed by~\cite{DWY_2016} and~\cite{DWY_2023} to solve the raking problem from Figure~\ref{fig:USHD_problem}. We suppose that there are $I$ different causes of deaths, $J$ different racial-ethnic groups, and $K$ counties in the state. We use $i = 0$ for the sum of deaths over all causes, $j = 0$ for the sum of deaths over all racial-ethnic groups, and $k = 0$ for the sum of deaths over all counties, i.e. the number of deaths at the state level. In the following, we denote
\begin{equation*}
f: \mathbb{R}^p \times \mathbb{R}^p \rightarrow \mathbb{R}^p
\end{equation*}
a distance function between the observations and the unknown raked values, $\mathbb{1}_p$ the unit vector in $\mathbb{R}^p$, $\odot$ the Hadamard product between two vectors:
\begin{equation*}
\left( a \odot b \right)_i = a_i b_i \quad \text{for} \quad a \in \mathbb{R}^p \quad , \quad b \in \mathbb{R}^p \quad \text{and} \quad a \odot b \in \mathbb{R}^p, 
\end{equation*}
and $\otimes$ the Kronecker product between two matrices:
\begin{equation*}
\left( A \otimes B \right)_{pr+v,qs+w} = a_{rs} b_{vw} \quad \text{for} \quad A \in \mathbb{R}^{m \times n} \quad , \quad B \in \mathbb{R}^{p \times q} \quad \text{and} \quad A \otimes B \in \mathbb{R}^{mp \times nq}
\end{equation*}

The raking problem from Figure~\ref{fig:USHD_problem} is decomposed in $2 \left( K + 1 \right)$ minimization problems.
\begin{enumerate}
    \item Compute the raked values, all-cause, all-racial-ethnic-group, by county. We need to ensure that the sum over counties of the all-cause, all-racial-ethnic-group deaths for each county is equal to the all-cause, all-racial-ethnic-group deaths for the state:
\begin{equation}
\label{eq:step1}
\min_{\beta_{0,0,.} \in \mathbb{R}^K} \mathbb{1}_K^T f \left( \beta_{0,0,.} , y_{0,0,.} \right) \quad \text{s.t} \quad \mathbb{1}_K^T \beta_{0,0,.} = s_0,
\end{equation}
where $y_{0,0,.} \in \mathbb{R}^K$ is the vector containing the number of all-cause, all-racial-ethnic-group deaths for each county, $\beta_{0,0,.}$ are the corresponding unknown raked values, and $s_0 \in \mathbb{R^+}$ is the total number of deaths for the state.

    \item Compute the raked values, all-racial-ethnic-group, by cause, by county. We need to ensure that, for each cause, the sums over counties of the all-racial-ethnic-group deaths for each county are equal to the all-racial-ethnic-group deaths for the state, and that, for each county, the sums over causes of the all-racial-ethnic-group deaths for each cause are equal to the all-cause, all-racial-ethnic-group deaths:
\begin{equation}
\label{eq:step2}
\min_{\beta_{.,0,.} \in \mathbb{R}^{IK}} \mathbb{1}_{IK}^T f \left( \beta_{.,0,.} , y_{.,0,.} \right) \quad \text{s.t.} \quad \left( \mathbb{1}_K \otimes I_I \right) \beta_{.,0,.} = s, \quad \left( I_K \otimes \mathbb{1}_I \right) \beta_{.,0,.} = \beta_{0,0,.},
\end{equation}
where $y_{.,0,.} \in \mathbb{R}^{IK}$ is the vector containing the number of all-racial-ethnic-group deaths for each cause and each county, $\beta_{.,0,.}$ are the corresponding unknown raked values, and $s \in \left( \mathbb{R}^+ \right)^I$ is the vector containing the number of deaths due to each cause for the state; 

    \item Compute the raked values, all-cause, by race, by county. For each county $k$, we need to ensure that the sum over racial-ethnic-groups of the all-cause deaths for each racial-ethnic-group for county $k$ is equal to the all-cause, all-racial-ethnic-group deaths for county $k$:
\begin{equation}
\label{eq:step3}
\min_{\beta_{0,.,k} \in \mathbb{R}^J} \mathbb{1}_J^T f \left( \beta_{0,.,k} , y_{0,.,k} \right) \quad \text{s.t.} \quad \mathbb{1}_J^T \beta_{0,.,k} = \beta_{0,0,k},
\end{equation}
where $y_{0,.,k} \in \mathbb{R}^J$ is the vector containing the number of all-cause deaths for each racial-ethnic-group for county $k$, $\beta_{.,0,k}$ are the corresponding unknown raked values, and $\beta_{0,0,k} \in \mathbb{R}$ is the raked value for all-cause, all-racial-ethnic-group deaths for county $k$.

    \item Compute the raked values, by cause, by racial-ethnic-group, by county. For each county $k$, we need to ensure that, for each cause, the sums over racial-ethnic-groups of the deaths for each racial-ethnic-group for county $k$ are equal to the all-racial-ethnic-group deaths for this county, and that, for each racial-ethnic-group, the sums over causes of the deaths for each cause for county $k$ are equal to the all-cause deaths for this county:
\begin{equation}
\label{eq:step4}
\min_{\beta_{.,.,k} \in \mathbb{R}^{IJ}} \mathbb{1}_{IJ}^T f \left( \beta_{.,.,k} , y_{.,.,k} \right) \quad \text{s.t.} \quad \left( \mathbb{1}_J \otimes I_I \right) \beta_{.,.,k} = \beta_{.,0,k}, \quad 
\left( I_J \otimes \mathbb{1}_I \right) \beta_{.,.,k} = \beta_{0,.,k},
\end{equation}
where $y_{.,.,k} \in \mathbb{R}^{IJ}$ is the vector containing the number of deaths for each cause, each racial-ethnic-group and county $k$, $\beta_{.,.,k}$ is the corresponding unknown raked values, $\beta_{.,0,k} \in \mathbb{R}^J$ is the vector containing the raked values for all-racial-ethnic-group deaths for each cause for county $k$, and $\beta_{0,.,k} \in \mathbb{R}^J$ is the vector containing the raked values for all-cause deaths for each racial-ethnic-group deaths for county $k$.

\end{enumerate}
If for $f$ we choose the entropic distance
\begin{equation}
\label{eq:entropic_distance}
f \left( \beta , y \right) = \beta \odot \log \left( \frac{\beta}{y} \right) - \beta + y \quad \text{for} \quad \beta , y \in \mathbb{R}^p
\end{equation}
then problem~\eqref{eq:step1} and the $K$ problems~\eqref{eq:step3} can be solved using the closed form solution (see Section 2 in Supplement 1)
\begin{equation*}
\beta_{0,0,.} = y_{0,0,.} \frac{s_0}{\mathbb{1}_K^T y_{0,0,.}} \quad \text \quad \beta_{0,.,k} = y_{0,.,k} \frac{\beta_{0,0,k}}{\mathbb{1}_J^T y_{0,.,k}} \text{ for } k = 1 , \cdots , K.
\end{equation*}
Problem~\eqref{eq:step2} and the $K$ problems~\eqref{eq:step4} can be solved using the Iterative Proportional Fitting algorithm (IFP,~\cite{DEM_1940},~\cite{STE_1942},~\cite{DEV_1992},~\cite{DEV_1993}, see Section 3 in Supplement 1).

In this paper, we replace the stagewise approach with $2 \left( K + 1 \right)$ minimization problems above by a single minimization problem. Denoting
\begin{equation*}
\begin{aligned}
& p = I J K, \\
& q = \left( I + 1 \right) \left( J + 1 \right) K, \\
& \beta = \left( \beta_{i,j,k} , i = 1 , \cdots , I , j = 1 , \cdots , J , k = 1 , \cdots , K \right) \in \mathbb{R}^p, \\
& w = \left( w_{i,j,k} , i = 0 , \cdots , I , j = 0 , \cdots , J , k = 1 , \cdots , K \right) \in \mathbb{R}^q, \\
& y = \left( y_{i,j,k} , i = 0 , \cdots , I , j = 0 , \cdots , J , k = 1 , \cdots , K \right) \in \mathbb{R}^q,
\end{aligned}
\end{equation*}
we formulate the more general problem
\begin{equation}
\label{eq:general_raking}
\min_{\beta \in \mathbb{R}^p} f^w \left( B \beta , y \right) \quad \text{s.t.} \quad A \beta = s,
\end{equation}
where
\begin{equation*}
A = \mathbb{1}_{JK}^T \otimes I_{I} \in \mathbb{R}^{I \times p},
\end{equation*}
\begin{equation*}
B = \begin{pmatrix} I_K \otimes \begin{pmatrix} \mathbb{1}_{IJ}^T \\ \mathbb{1}_J^T \otimes I_I \\ I_J \otimes \mathbb{1}_I^T \end{pmatrix} \\ I_{IJK} \end{pmatrix} \in \mathbb{R}^{q \times p}
\end{equation*}
and
\begin{equation*}
f^w \left( \zeta , y \right) = \sum_{i = 1}^q w_i f_i \left( \zeta_i , y_i \right) \quad \text{for} \quad w , \zeta , y \in \mathbb{R}^q.
\end{equation*}
In~\eqref{eq:general_raking}, the matrix $A$ encodes the aggregates over racial-ethnic-groups and counties that \textbf{must match} the known margins $s$, which are the number of deaths for each cause at the state level. The matrix $B$ is a stack of the remaining aggregate operations that should be consistent (but not necessarily exactly equal to) marginal observations (see steps~\eqref{eq:step2},~\eqref{eq:step3} and~\eqref{eq:step4}), as well as the identity matrix which limits how much raked values deviate from initial detailed observations.

To solve~\eqref{eq:general_raking}, we leverage convex duality, since the dual problem lends itself to efficient solution methods. We introduce an auxiliary variable $\zeta \in \mathbb{R}^q$ with constraint $\zeta := B \beta$, and, using the convex preliminaries in Section 1 of Supplement 2, we summarize the original problem~\eqref{eq:general_raking},  Lagrangian associated to the equality constraints, resulting dual problem, and primal dual relationships that allow us to recover the primal solution once the dual is solved:
\begin{eqnarray*}
\mathcal{P}: & \quad & \min_{\beta \in \mathbb{R}^p , \zeta \in \mathbb{R}^q} f^w \left( \zeta , y \right) \quad \text{s.t.} \quad \begin{pmatrix} A & 0 \\ B & -I \end{pmatrix} \begin{pmatrix} \beta \\ \zeta \end{pmatrix} = \begin{pmatrix} s \\ 0 \end{pmatrix} \\
\mathcal{L}: & \quad & f^w \left( \zeta , y \right) + \lambda_A^T \left( A \beta - s \right) + \lambda_B^T \left( B \beta - \zeta \right) \\
\mathcal{D}: & \quad & \min_{\lambda_A \in \mathbb{R}^I , \lambda_B \in \mathbb{R}^q} \lambda_A^T s + f^{w*} \left( \lambda_B , y \right) + \delta_0 \left( - A^T \lambda_A - B^T \lambda_B \right). \\
\text{OPT}: & \quad & \zeta^* = \nabla_z f^{w*} \left( \lambda_B^* , y \right) , \quad \begin{pmatrix} A \\ B \end{pmatrix} \beta^* = \begin{pmatrix} s \\ \zeta^* \end{pmatrix}
\end{eqnarray*}
where $f^{w*}: \mathbb{R}^q \rightarrow \mathbb{R}$ is the convex conjugate of $f^w$

\begin{equation*}
f^{w*} \left( z \right) = \sup_{\zeta \in \mathbb{R}^q} z^T \zeta - f^w \left( \zeta \right) \quad \text{for} \quad z \in \mathbb{R}^q,
\end{equation*}
$\delta_0$ is the indicator function
\begin{equation*}
\delta_0 \left( x \right) = \begin{cases}
0 & x = 0 \\
\infty & x \neq 0
\end{cases}
\end{equation*}
and $\nabla_z f^{w*}: \mathbb{R}^q \rightarrow \mathbb{R}^q$ is the gradient of $f^{w*}$.

To reduce the dimensions of problem~\eqref{eq:general_raking} and at the same time rigorously check for consistency of constraints (guaranteeing the existence of a solution), we can rewrite the matrix $B$ as
\begin{equation*}
B = \begin{pmatrix} B_1 \\ I_p \end{pmatrix} \quad \text{with} \quad B_1 = I_K \otimes \begin{pmatrix} \mathbb{1}_{IJ}^T \\ \mathbb{1}_J^T \otimes I_I \\ I_J \otimes \mathbb{1}_I^T \end{pmatrix} \in \mathbb{R}^{\left( q -  p \right) \times p}
\end{equation*}
The equality constraints in the primal formulation of~\eqref{eq:general_raking} can then be rewritten as
\begin{equation*}
\begin{pmatrix} 0_{k,q-p} & A \\ - I_{q-p} & B_1 \end{pmatrix} \zeta = \begin{pmatrix} s \\ 0_{q-p} \end{pmatrix}
\end{equation*}
Denoting
\begin{equation*}
A^* = \begin{pmatrix} 0_{k,q-p} & A \\ - I_{q-p} & B_1 \end{pmatrix} \quad \text{and} \quad s^* = \begin{pmatrix} s \\ 0_{q-p} \end{pmatrix}
\end{equation*}
we get the reduced primal formulation, Lagrangian, dual formulation, and primal-dual relationships:
\begin{equation}
\label{eq:simplified_general}
\begin{aligned}
\mathcal{P}: & \quad \min_{\zeta \in \mathbb{R}^q} f^w \left( \zeta , y \right) \quad \text{s.t.} \quad A^* \zeta = s^* \\
\mathcal{L}: & \quad f^w \left( \zeta , y \right) + \lambda^T \left( A^* \zeta - s^* \right) \\
\mathcal{D}: & \quad \min_{\lambda \in \mathbb{R}^{I + q - p}} \lambda^T s^* + f^{w*} \left( - A^{*T} \lambda , y \right). \\
\text{OPT}: & \quad \zeta^* = \nabla_z f^{w*} \left( - A^{*T} \lambda^* , y \right) \quad \text{and} \quad \beta^* = \begin{pmatrix} 0_{p,q-p} & I_p \end{pmatrix} \zeta^*
\end{aligned}
\end{equation}

We have thus reduced the size of the problem from $q$ to $I + q - p << q$. The primal problem will have a unique solution $\zeta^*$ if $\text{rank}\left( A^* \right) = I + q - p $. Using the structure of the matrix $A^*$ above, we can easily see that this is the case.

\subsection{Variance propagation}
\label{sec:uncertainty}

Calculating the variance of the estimator of the raked values is important for all variants of the raking problem. In the survey sampling domain, we are generally interested in estimating a raked value for the sum over all categorical variables, which would be in our case $\beta_{0,0,0}^*$, the number of deaths from all causes, for all races, at the state level. The observations $y$ are known without uncertainty, but represent a sample $S$ selected from the total population $\mathcal{U}$. Therefore, the variance of the estimator is informed by the selection of the sample $S$ from the population $\mathcal{U}$ and the uncertainty on the margins $s$. The Horvitz-Thompson-Narain estimator with accompanying variance was first proposed by~\cite{NAR_1951} and~\cite{HOR_1952}. ~\cite{LU_2003} used the uncertainty on the raking weights to compute the variance of the resulting estimate. ~\cite{KIM_2011} accounted for the raking procedure to estimate the variance of the estimator. ~\cite{MOR_2021} used a bootstrap method to estimate the uncertainty of the raking estimator for small area estimation. A review of methods to use priors on the margins to compute the posterior distribution for the total is given by~\cite{SI_2021}.

In the context of global health, the sample $S$ would be equal to the whole population $\mathcal{U}$ and is not associated with any uncertainty. However, the observations $y$ are uncertain and we are interested in using the known variances and covariances of the observations $y$ and the margins $s$ to estimate the unknown covariance matrix of the raked values $\beta^*$ and their aggregates $\zeta^*$.

\subsubsection{Computation of the covariance matrix}
\label{sec:covariance}

In the previous work from~\cite{DWY_2016} and ~\cite{DWY_2023}, $n$ samples $\left( y^{\left( i \right)} , s^{\left( i \right)} \right) , i = 1 , \cdots n$ from the random vector $\left( y , s \right)$ were available. The authors applied the raking procedure from Section ~\ref{sec:raking} to each of the samples to compute the corresponding raked values $\beta^{* \left( i \right)}$. They then computed the sample covariance matrix to get the variance of the estimator $\beta^*$. This procedure comes at a high storage and computational cost as all the samples are first generated from a known covariance matrix and must be stored in memory and the raking procedure must then be applied to each one of them. In this paper, we propose to use only the sample mean $\left( \bar{y} , \bar{s} \right)$ of the observations and margins, their covariance matrix $\Sigma$ and a single raking procedure to compute the covariance matrix of the raked values $\beta^*$ and their aggregates $\zeta^*$. Let us consider the minimization problem
\begin{equation*}
\mathcal{P}: \quad \min_{\beta \in \mathbb{R}^p, \zeta \in \mathbb{R}^q} f^w \left( \zeta , y \right) \quad \text{s.t.} \quad \begin{pmatrix} A & 0_{I \times q} \\ B & - I_q \end{pmatrix} \begin{pmatrix} \beta \\ \zeta \end{pmatrix} = \begin{pmatrix} s \\ 0_q \end{pmatrix}
\end{equation*}
with Lagrangian
\begin{equation*}
\mathcal{L} \left( \beta , \zeta , \lambda_A , \lambda_B \right): \quad f^w \left( \zeta , y \right) + \lambda_A^T \left( A \beta - s \right) + \lambda_B^T \left( B \beta - \zeta \right).
\end{equation*}
We define
\begin{equation*}
F \left( \beta^* , \zeta^* , \lambda_A^* , \lambda_B^* ; y , s \right) := \nabla \mathcal{L}\left( \beta^* , \zeta^* , \lambda_A^* , \lambda_B^* \right) = \begin{bmatrix} A^T \lambda_A^* + B^T \lambda_B^* \\ \nabla_\zeta f^w \left( \zeta^* ; y \right) - \lambda_B^* \\ A \beta^* - s \\ B \beta^* - \zeta^* \end{bmatrix} = 0.
\end{equation*}
The solution of this system can be written as
\begin{equation*}
\begin{pmatrix} \beta^* \\ \zeta^* \end{pmatrix} = \phi \left( y ; s \right) \text{ with } \phi : \mathbb{R}^{q + I} \rightarrow \mathbb{R}^{p + q}.
\end{equation*}
\begin{theorem}\label{th1}
Suppose that the vector of observations and margins $\left( y , s \right)$ is a random vector with expectation $\theta$ and covariance matrix $\Sigma$. Given $n$ samples, denoted $\left( y^{\left( i \right)}, s^{\left( i \right)}\right)$ for $i = 1, \cdots, n$, denote their sample mean by $\left( \bar{y}_n, \bar{s}_n \right)$. Let $\phi'_\theta$ denote the matrix of the partial derivatives of $\phi$ with respect to the observations $y$ and margins $s$ taken at $\theta$. If $\begin{pmatrix} \beta^*_n \\ \zeta^*_n \end{pmatrix}$ denotes the raked values of the sample mean and their aggregates, then the covariance matrix of the raked values and their aggregates $\begin{pmatrix} \beta^*_n \\ \zeta^*_n \end{pmatrix}$ is given by
\begin{equation*}
\Sigma_{\beta_n , \zeta_n} = \phi'_\theta \Sigma \phi^{'T}_\theta.
\end{equation*}
\end{theorem}
The proof is given in Section 2 of Supplement 2.

To understand this process better, consider the raked values of the random vector $\left( y, s \right)$, denoted $\phi \left( y , s \right)$ with expectation $\mathbb{E} \phi$ and covariance matrix $\Sigma_{\beta , \zeta}$. Given $n$ samples of the raked vector $\phi \left( y^{\left( i \right)}, s^{\left( i \right)} \right)$, from the central limit theorem we have
\begin{equation}
\label{eq:CLT}
\sqrt{n} \left( \frac{1}{n} \sum_{i = 0}^n \phi \begin{pmatrix} y^{\left( i \right)} \\ s^{\left( i \right)} \end{pmatrix} - \mathbb{E} \phi \begin{pmatrix} y \\ s \end{pmatrix} \right) \rightarrow \mathcal{N} \left( 0 , \Sigma_{\beta , \zeta} \right).
\end{equation}
If to first order we have 
\begin{equation}
\label{eq:requirement}
\frac{1}{n} \sum_{i = 0}^n \phi \begin{pmatrix} y^{\left( i \right)} \\ s^{\left( i \right)} \end{pmatrix} \approx \phi \begin{pmatrix} \bar{y}_n \\ \bar{s}_n \end{pmatrix},
\end{equation}
then from Theorem \ref{th1} and~\eqref{eq:CLT} the draws process has the asymptotically correct statistics
\begin{equation*}
\mathbb{E} \phi \begin{pmatrix} y \\ s \end{pmatrix} \approx \phi \left( \theta \right) \text{ and } \Sigma_{\beta , \zeta} \approx \phi'_\theta \Sigma \phi^{'T}_\theta.
\end{equation*}
The approximation~\eqref{eq:requirement} is an equality when $\phi$ is a linear map, for example, in linear regression.  When $\phi$ is nonlinear, or affine constraints are present,~\eqref{eq:requirement} may not hold. For the raking problem, we show empirically in the experiments that given enough draws we actually can get close to the asymptotic statistics, but we incur a systematic bias when we use the draws. 

\subsubsection{Implicit Function Theorem and derivative computations}
\label{sec:IFT}

To compute $\Sigma_{\beta , \zeta}$, we now need to compute the partial derivatives $\frac{\partial \phi_i}{\partial y_j}$ and $\frac{\partial \phi_i}{\partial s_j}$. We adapt the statement of the Implicit Function Theorem from~\cite{FOL_2002}. Assume that $S$ is an open subset of $\mathbb{R}^{p + 3q + 2I}$ and that $F : S \rightarrow \mathbb{R}^{p + q}$ is a function of class $C^1$. Assume also that $\left( \beta^* , \zeta^* , \lambda_A^* , \lambda_B^* ; y_0 , s_0 \right)$ is a point in $S$ such that
\begin{equation*}
F \left( \beta^* , \zeta^* , \lambda_A^* , \lambda_B^* ; y_0 , s_0 \right) = 0, \quad  \det D_{\beta , \zeta , \lambda_A , \lambda_B} F \left( \beta^* , \zeta^* , \lambda_A^* , \lambda_B^* ; y_0 , s_0 \right) \neq 0
\end{equation*}
where $D_{\beta , \zeta , \lambda_A , \lambda_B} F$ is the matrix of partial derivatives.
We have the following:
\begin{enumerate}
\item[i.] There exist $r_0 , r_1 > 0$ such that for every $\left( y , s \right) \in \mathbb{R}^{q + I}$ with
\begin{equation*}
\left\lVert \left( y , s \right) - \left( y_0 , s_0 \right) \right\rVert < r_0,
\end{equation*}
there is a unique $\left( \beta , \zeta , \lambda_A , \lambda_B \right) \in \mathbb{R}^{p + 2q + I}$ such that
\begin{equation}
\label{eq:defIFT}
\left\lVert \left( \beta , \zeta , \lambda_A , \lambda_B \right) - \left( \beta^* , \zeta^* , \lambda_A^* , \lambda_B^* \right) \right\rVert < r_1 \text{ and } F \left( \beta , \zeta , \lambda_A , \lambda_B ; y , s \right) = 0.
\end{equation}
Thus~\eqref{eq:defIFT} implicitly defines a function $\left( \beta , \zeta , \lambda_A , \lambda_B \right) = \phi \left( y , s \right)$ for $\left( y , s \right) \in \mathbb{R}^{q + I}$ near $\left( y_0 , s_0 \right)$, with $\left( \beta , \zeta, \lambda_A , \lambda_B \right) = \phi \left( y , s \right)$ close to $\left( \beta^* , \zeta^* , \lambda_A^* , \lambda_B^* \right)$. In particular $\phi \left( y_0 , s_0 \right) = \left( \beta^* , \zeta^* , \lambda_A^* , \lambda_B^* \right)$.

\item[ii.] Moreover, the function $\phi : B \left( r _0 ; y , s \right) \rightarrow B \left( r_1 ; \beta , \zeta , \lambda_A , \lambda_B \right) \subset \mathbb{R}^{p + 2q + I}$ above is continuously differentiable, and its derivatives may be determined by 
differentiating $F \left( y , s ; \phi \left( y , s \right) \right) = 0$ at the solution $\left( \beta^* , \zeta^* , \lambda_A^* , \lambda_B^* \right)$ to obtain
\begin{equation}
\label{eq:uncertaintySyst}
\left[ D_{\beta , \zeta , \lambda_A , \lambda_B} F \left( y , s ; \beta^* , \zeta^* , \lambda_A^* , \lambda_B^* \right) \right] \left[ D_{y , s} \phi \left( y , s \right) \right] = - \left[ D_{y , s} F \left( y , s ; \beta^* , \zeta^* , \lambda_A^* , \lambda_B^* \right) \right].
\end{equation}
\end{enumerate}
We can compute the matrices of partial derivatives
\begin{equation*}
D_{\beta , \zeta , \lambda_A , \lambda_B} F = \begin{pmatrix} 0_{p \times p} & 0_{p \times q} & A^T & B^T \\ 0_{q \times p} & \nabla^2_\zeta f^w \left( \zeta^* ; y \right) & 0_{q \times I} & -I_q \\ A & 0_{I \times q} & 0_{I \times I} & 0_{I \times q} \\ B & -I_q & 0_{q \times I} & 0_{q \times q} \end{pmatrix}
\end{equation*}
and
\begin{equation*}
D_{y , s} F = \begin{pmatrix} 0_{p \times q} & 0_{p \times I} \\\nabla^2_{\zeta y} f^w \left( \zeta^* ; y \right) & 0_{q \times I} \\ 0_{I \times q} & - I_{I \times I} \\ 0_{q \times q} & 0_{q \times I} \end{pmatrix}.
\end{equation*}
For separable distance functions $f^w$ as in all of our examples, the Hessian matrices $\nabla^2_\zeta f^w \left( \zeta ; y \right)$ and $\nabla^2_{\zeta y} f^w \left( \zeta ; y \right)$ are diagonal. We have chosen the constraints such that $\text{rank} \begin{pmatrix} A \\ B \end{pmatrix} = p$,  so that $D_{\beta , \zeta , \lambda_A , \lambda_B} F$ is invertible, and we solve system~\eqref{eq:uncertaintySyst} for $D_{y,s}\phi \left( y , s \right)$. 

\subsubsection{Synthetic example}
\label{synthetic_example}

To illustrate the uncertainty quantification method explained in Section~\ref{sec:covariance} and Section~\ref{sec:IFT}, we apply the raking procedure to a small synthetic 2-dimensional example with two categorical variables $X_1$ and $X_2$. We suppose that $X_1$ can take $m = 3$ values $\left[ 1 ; 2 ; 3 \right]$ and that $X_2$ can take $n = 5$ values $\left[ 1 ; 2 ; 3 ; 4 ; 5 \right]$ and we generate $m \times n$ values following a uniform distribution over the interval $\left[ 2 ; 3 \right]$. We compute the margins $s_r = \beta_0 \mathbb{1}_n$ and $s_c = \beta_0^T \mathbb{1}_m$. We then add noise to the balanced table to obtain an unbalanced table $y_0 = \beta_0 + \mathcal{N} \left( 0 , 0.1 \right)$. We generate $N$ samples following a multivariate normal distribution with expectancy $y_0$ and covariance $\Sigma$, with off-diagonal elements equal to $0.01$ and  diagonal elements set to $\Sigma_{k,k} = 0.1 \times k$ for $k = 1 , \cdots , m \times n$. We then apply the raking procedure to the expectancy to get $\beta^* = \phi \left( y_0 , s_1 , s_2 \right)$. 

\begin{figure}[h!]
\begin{center}
\includegraphics[width=\textwidth]{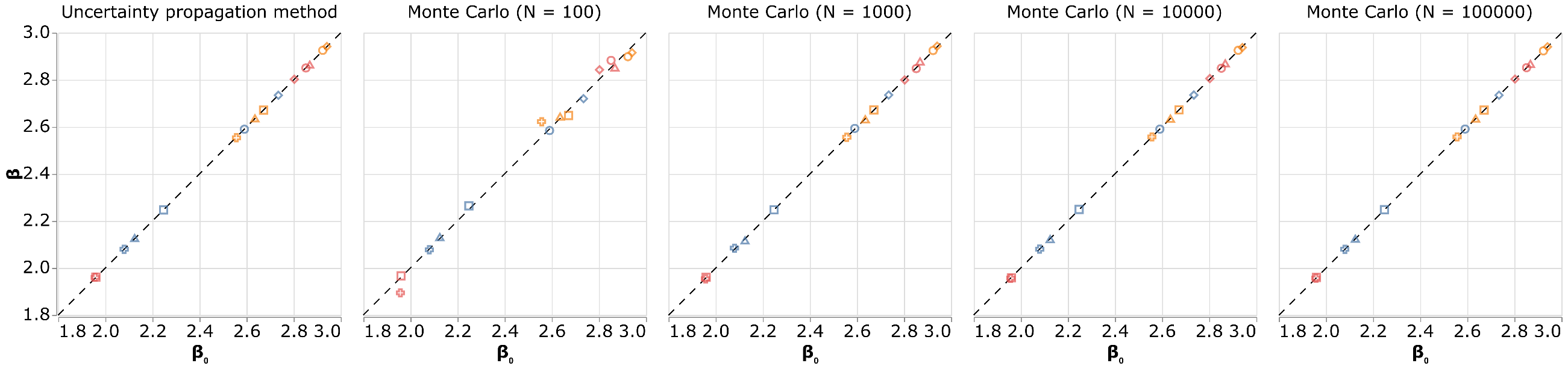}
\end{center}
\caption{Estimator of the raked values computed with the proposed uncertainty propagation method (left) or by taking the sample mean of the raked values with $N = 100, 1000 , 10000, 100000$ samples (right). The proposed uncertainty propagation method gives better results than the Monte Carlo method when less than $1000$ samples are available. \label{fig:synthetics_mean}}
\end{figure}

\begin{figure}[h!]
\begin{center}
\includegraphics[width=\textwidth]{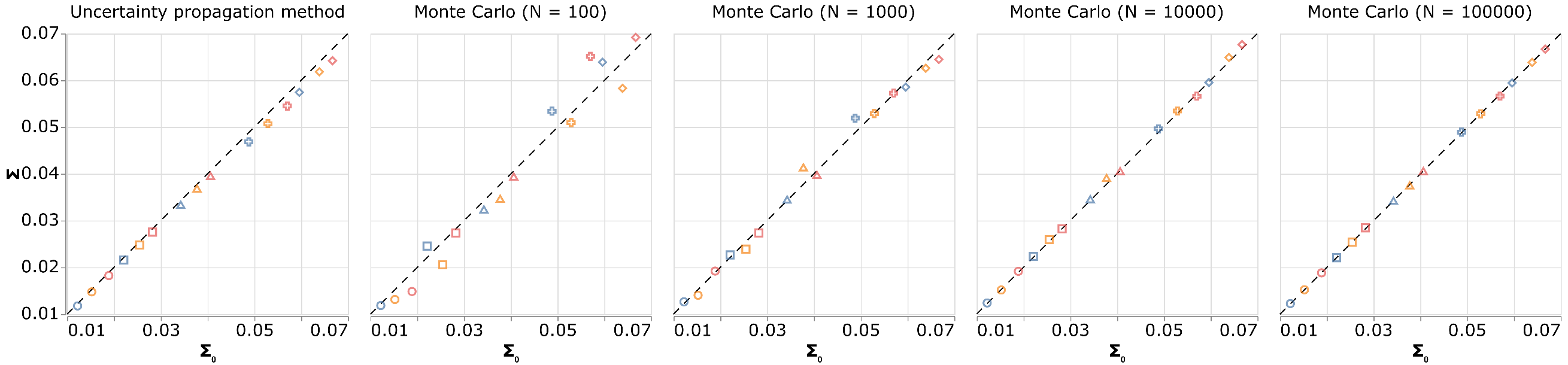}
\end{center}
\caption{Estimator of the variance of the raked values computed with the proposed uncertainty propagation method (left) or by taking the sample variance of the raked values with $N = 100, 1000 , 10000, 100000$ (right). The proposed uncertainty propagation method is close to the Monte Carlo method when the latter uses many samples. More variation is evident in the Monte Carlo results for fewer samples.}
\label{fig:synthetics_variance}
\end{figure}

We compare our uncertainty propagation method to a Monte Carlo method to estimate the expectancies and variances of the raked values. For the gold standard, we compute the sample mean and the sample variance of the raked values using $10^6$ samples. We then compare in Figure~\ref{fig:synthetics_mean} the raked value $\beta^* = \phi \left( y_0 , s_1 , s_2 \right)$ obtained when raking the expectancy $y_0$ of the observations and the sample mean of the raked values $\bar{\beta} = \frac{1}{N} \sum_{i = 1}^N \phi \left( y^{\left( i \right)} , s_1 , s_2 \right)$ when raking each of the samples $y^{\left( i \right)}$, for $N = 100, 1000 , 10000 , 100000$. We need at least $1000$ samples to obtain the same result when raking the samples as when raking the expectancy.
We also compare in Figure~\ref{fig:synthetics_variance} the variance $\text{diag} \left( \Sigma_\beta \right) = \text{diag} \left( \phi_{y_0}' \Sigma \phi_{y_0}^{'T} \right)$ of the raked values obtained when raking the expectancy $y_0$ of the observations and using the uncertainty propagation method and when raking each of the samples $y^{\left( i \right)}$ and computing the sample variance of the raked values $\text{diag} \left( S \right) = \text{diag} \left( \frac{1}{N} \sum_{i = 1}^N \left( \phi \left ( y^{\left( i \right)} , s_1 , s_2 \right) - \bar{\beta} \right)^2 \right)$, for $N = 100, 1000 , 10000 , 100000$. We need at least $10000$ samples to obtain better results when raking the samples than when using the uncertainty propagation method. Our method still has a slight bias compared to the reference. However, in practice we do not expect to have more than a few hundred samples available. Despite the bias, the uncertainty propagation method will then give more accurate results than the Monte Carlo method.

\section{Raking extensions}
\label{sec:discussion}

The general raking method described in Section ~\ref{sec:methods} has additional advantages for the user, compared to the initial method from~\cite{DWY_2016} and ~\cite{DWY_2023}.

\subsection{Raking with differential weights}
\label{sec:method_weights}

Observations $y$ are modeled quantities that come with uncertainty information. Introducing weights allows us to account for these uncertainties so that more certain inputs move less during the raking process compared to less certain inputs. For example, a simple approach to reweighting is to take the weights $w_i$ equal to $\frac{1}{\sigma_i^2}$ where $\sigma_i$ is the standard deviation of the reported $y_i$. 

From the definition of the weighted loss function
\begin{equation*}
f^w \left( \zeta , y \right) = \sum_{i = 1}^q w_i f_i \left( \zeta_i , y_i \right),
\end{equation*}
we have
\begin{equation*}
\left( f_i^{w_i} \right)^* \left( z_i \right) = w_i f_i^* \left( \frac{z_i}{w_i}\right), \quad \zeta^* = \nabla_z f^*\left( \lambda_B^* \odot \frac{1}{w} , y \right). 
\end{equation*}
As any weight $w_i$ goes to infinity, the corresponding solution $\zeta_i^*$ goes to $\nabla_z f_i^* \left( 0 \right)$. If we use for $f$ the entropic distance~\eqref{eq:entropic_distance}, this gives us $\zeta_i^* = y_i$, that is, the observations with no uncertainty are kept constant through the raking process. We illustrate in Section 1 in Supplement 3 how raking with weights allows retrieving the true value of a biased observation.

\subsection{Raking with different losses}
\label{sec:raking_losses}

The entropic distance~\eqref{eq:entropic_distance} appears to be a natural choice of raking objective given the solutions of problems~\eqref{eq:step1},~\eqref{eq:step2},~\eqref{eq:step3} and~\eqref{eq:step4}. However, simpler functions have computational advantages,  while more complex functions achieve additional aims, such as maintaining upper and lower bounds. In this section, we present alternative raking losses and summarize all for convenience. All of the functions we use here are positive, convex and differentiable, with differentiable conjugates; they also satisfy $f \left( y , y \right) = 0$. 

\subsubsection{Weighted least squares}

We can compute the Taylor expansion of the entropic distance~\eqref{eq:entropic_distance} centered at $x = y$:
\begin{eqnarray*}
& f \left( \beta ; y \right) = \beta \log \left( \frac{\beta}{y} \right) - \left( \beta -y \right) \quad & \Rightarrow \quad f \left( y , y \right) = 0 \\
& f' \left( \beta ; y \right) = \log \left( \frac{\beta}{y} \right) \quad & \Rightarrow \quad f' \left( y , y \right) = 0 \\
& f'' \left( \beta ; y \right) = \frac{1}{\beta} \quad & \Rightarrow \quad f'' \left( y , y \right) = \frac{1}{y}
\end{eqnarray*}
Thus the second order Taylor expansion of the entropic distance function~\eqref{eq:entropic_distance} gives rise to a weighted least squares, or $\chi$-square:
\begin{equation*}
\beta \log \left( \frac{\beta}{y} \right) - \left( \beta - y \right) \approx \frac{1}{2y} \left( \beta - y \right)^2 := f_2 \left( \beta ; y \right). 
\end{equation*}
If we take $B = I_p$ and $w = \mathbb{1}_p$, the dual of the problem~\eqref{eq:general_raking} simplifies into
\begin{equation*}
\mathcal{D}: \quad \min_{\lambda_A \in \mathbb{R}^I} \lambda_A^T s + f_2^* \left( - A^T \lambda_A \right).
\end{equation*}
Using the conjugate of $f_2$, $f_2^* \left( z , y \right) = y^T \left( \frac{z^2}{2} + z \right)$, the dual problem becomes:
\begin{equation*}
\min_{\lambda_A \in \mathbb{R}^I} \left\{ \frac{1}{2} \lambda_A^T A Y A^T \lambda_A - \lambda_A^T \left( A y - s \right) \right\},
\end{equation*}
with solution:
\begin{equation*}
\lambda_A^* = \left( A Y A^T \right)^{-1} \left( A y - s \right) \quad \text{and} \quad \beta^* = y \odot \left( 1 - A^T \left( A Y A^T \right)^{-1} \left( A y - s \right) \right). 
\end{equation*}
In this case, the reader can explicitly check that $A\beta^* = s$. The closed form rake is convenient, but the output is not guaranteed to be positive when the input is positive. 

\subsubsection{Entropic loss} The convex conjugate of the entropic loss~\eqref{eq:entropic_distance} is
\begin{equation*}
f^* \left( z , y \right) = y^T \left( \exp \left( z \right) - 1 \right),
\end{equation*}
and thus the solution of problem~\eqref{eq:general_raking} becomes
\begin{equation*}
\zeta^* = y \odot \exp \left( \frac{1}{w} \odot \lambda_B^* \right).
\end{equation*}
The dual solution does not admit a closed form, but the primal-dual relationship makes it clear that the aggregates of the raked values are obtained by differential scaling, and outputs are positive when the inputs are positive and zero when inputs are zero.  

\subsubsection{Logistic loss} In our application case, the observations are bounded quantities, as we expect the death counts to be positive and lower than the population counts; thus, we need lower and upper bounds on each aggregated raked value in addition to the aggregation constraints. We can seamlessly add bound constraints by adopting a logistic loss with domain $l \leq \zeta \leq u$:
\begin{equation*}
f_i \left( \zeta_i, y_i \right) = \left( \zeta_i - l_i \right) \log \left( \frac{\zeta_i - l_i}{y_i - l_i} \right) + \left( u_i - \zeta_i \right) \log \left( \frac{u_i - \zeta_i}{u_i - y_i} \right).
\end{equation*}
The calculation of the final conjugate is detailed in Section 3 in Supplement 2. We have
\begin{equation*}
f^* \left( z \right) = \mathbb{1}_p^T \left( \left( u - l \right) \odot \log \left( \frac{y - l}{u - l} \odot \exp \left( z \right) + \frac{u - y}{u - l} \right) + l \odot z \right),
\end{equation*}
which finally gives us
\begin{equation*}
\zeta^* = l + \left( u - l \right) \odot \left( \frac{ \left( y - l \right) \odot \exp \left( \lambda_B^* \odot \frac{1}{w} \right)}{ \left( y -l \right) \odot \exp \left( \lambda_B^* \odot \frac{1}{w} \right) + u - y} \right),
\end{equation*}
which the reader can verify satisfies the bound constraint $l \leq \zeta^* \leq u$ whenever $y$ satisfies the same constraint. The logistic loss thus enforces the bounds while preserving the simple problem structure of~\eqref{eq:general_raking}. 

We show in Section 2 in Supplement 3 how the choice of the raking loss affects the final raked values.

\subsection{Feasibility of constraints}

The objective functions in Section~\ref{sec:raking_losses} are only defined when $y \neq 0$ for the weighted least squares and the entropic loss, or when $y \neq l $ and $y \neq u$ for the logistic loss. In the former cases, we want to respect the additional constraints $\zeta_i = 0$ wherever the corresponding observations $y_i = 0$. In the later case, we want to respect the additional constraints $\zeta_i = l_i$ wherever the corresponding observations $y_i = l_i$ and $\zeta_i = u_i$ wherever the corresponding observations $y_i = u_i$.

Let us consider the case of the weighted least squares and the entropic loss. We consider the matrix $P$ to be a permutation matrix that select the $\tilde{q} \leq q$ entries of $y$ that are non-zeros and $Q$ to be a permutation matrix that select the $q - \tilde{q}$ entries of $y$ that are equal to $0$. We thus have
\begin{equation*}
P^T P + Q^T Q = I_q
\end{equation*}
The problem~\eqref{eq:simplified_general} then becomes
\begin{equation*}
\begin{aligned}
& \min_{\zeta \in \mathbb{R}^q} f^w \left( P \zeta , P y \right) \quad \text{s.t.} \quad A^* \zeta = s^* \quad \text{and} \quad Q \zeta = 0 \\
& \min_{\zeta \in \mathbb{R}^q} f^w \left( P \zeta , P y \right) \quad \text{s.t.} \quad A^* \left( P^T P + Q^T Q \right) \zeta = s^* \quad \text{and} \quad Q \zeta = 0 \\
& \min_{\tilde{\zeta} \in \mathbb{R}^{\tilde{q}}} f^w \left( \tilde{\zeta} , \tilde{y} \right) \quad \text{s.t.} \quad A^* P^T \tilde{\zeta} = s^* \quad \text{and} \quad Q \zeta = 0.
\end{aligned}
\end{equation*}
We thus only need to solve the reduced problem
\begin{equation}
\label{eq:reduced_raking}
\begin{aligned}
\mathcal{P}: & \quad \min_{\tilde{\zeta} \in \mathbb{R}^{\tilde{q}}} f^w \left( \tilde{\zeta} , \tilde{y} \right) \quad \text{s.t.} \quad A^* P^T \tilde{\zeta} = s^* \\
\mathcal{L}: & \quad f^w \left( \tilde{\zeta} , \tilde{y} \right) + \tilde{\lambda}^T \left( A^* P^T \tilde{\zeta} - s^* \right) \\
\mathcal{D}: & \quad \min_{\tilde{\lambda} \in \mathbb{R}^{I + q - p}} \tilde{\lambda}^T s^* + f^{w*} \left( - P A^{*T} \lambda , \tilde{y} \right). \\
\text{OPT}: & \quad \tilde{\zeta}^* = \nabla_z f^{w*} \left( - P A^{*T} \tilde{\lambda}^* , \tilde{y} \right)
\end{aligned}
\end{equation}

$\text{rank} \left( A^* P^T \right) = I + q - p$ is thus a sufficient condition for the raking problem to have a solution. However, depending on which observations $y_i$ are zeros, we may have cases where $\text{rank} \left( A^* P^T \right) < I + q - p$ and the problem does not have a solution.

We now consider the case of the logistic loss. We consider the matrix $P$ to be a permutation matrix that select the $\tilde{q} \leq q$ entries of $y$ that are different from $l$ and $u$, $Q$ to be a permutation matrix that select the entries of $y$ that are equal to $l$ and $R$ to be a permutation matrix that select the entries of $y$ that are equal to $u$. We thus have
\begin{equation*}
P^T P + Q^T Q + R^T R = I_q
\end{equation*}
The problem~\eqref{eq:simplified_general} then becomes
\begin{equation*}
\begin{aligned}
& \min_{\zeta \in \mathbb{R}^q} f^w \left( P \zeta , P y \right) \quad \text{s.t.} \quad A^* \zeta = s^* , \quad Q \zeta = Q l \quad \text{and} \quad R \zeta = R u \\
& \min_{\zeta \in \mathbb{R}^q} f^w \left( P \zeta , P y \right) \quad \text{s.t.} \quad A^* \left( P^T P + Q^T Q + R^T R \right) \zeta = s^* , \quad Q \zeta = Q l \quad \text{and} \quad R \zeta = R u \\
& \min_{\tilde{\zeta} \in \mathbb{R}^{\tilde{q}}} f^w \left( \tilde{\zeta} , \tilde{y} \right) \quad \text{s.t.} \quad A^* P^T \tilde{\zeta} = s^* - A^* Q^T Q l - A^* R^T R u , \quad Q \zeta = Q l \quad \text{and} \quad R \zeta = R u.
\end{aligned}
\end{equation*}
We thus only need to solve the reduced problem
\begin{equation}
\label{eq:reduced_raking_logit}
\begin{aligned}
\mathcal{P}: & \quad \min_{\tilde{\zeta} \in \mathbb{R}^{\tilde{q}}} f^w \left( \tilde{\zeta} , \tilde{y} \right) \quad \text{s.t.} \quad A^* P^T \tilde{\zeta} = s^* - A^* Q^T Q l - A^* R^T R u \\
\mathcal{L}: & \quad f^w \left( \tilde{\zeta} , \tilde{y} \right) + \tilde{\lambda}^T \left( A^* P^T \tilde{\zeta} - s^* + A^* Q^T Q l + A^* R^T R u\right) \\
\mathcal{D}: & \quad \min_{\tilde{\lambda} \in \mathbb{R}^{I + q - p}} \tilde{\lambda}^T \left( s^* - A^* Q^T Q l - A^* R^T R u \right) + f^{w*} \left( - P A^{*T} \lambda , \tilde{y} \right). \\
\text{OPT}: & \quad \tilde{\zeta}^* = \nabla_z f^{w*} \left( - P A^{*T} \tilde{\lambda}^* , \tilde{y} \right)
\end{aligned}
\end{equation}

Here again, $\text{rank} \left( A^* P^T \right)  = I + q - p$ or $s^* - A^* Q^T Q l - A^* R^T R u$ is in the column space of $A^* P^T$ is a sufficient condition for the problem to have a solution.

\subsection{Raking with missing data}
\label{sec:method_missing}

The framework from problem~\eqref{eq:general_raking} can also be used to impute raked values even if the corresponding observations are missing from the initial table $y$. Let us decompose the matrix $B$ into two parts: $B = \begin{pmatrix} P \\ C \end{pmatrix}$, where 
$P$ is a permutation matrix that selects the $\widetilde q \leq q$ entries of $y$ that have observations and $C$ is a matrix that computes the aggregates. We can define
\begin{equation*}
\lambda_B = \begin{bmatrix} \lambda_P \\ \lambda_C \end{bmatrix} \quad \text{and} \quad \zeta := P \beta + C \beta
\end{equation*}
and we retrieve the same structure as the initial problem~\eqref{eq:general_raking}. However, there is more structure that we can exploit within the affine constraint in $\mathcal{D}$ and we can reduce the problem dimension and complexity. The dual affine constraint $A^T \lambda_A + P^T \lambda_P + C^T \lambda_C$ can be used to solve for $\lambda_P$:
\begin{equation*}
\lambda_P = -P \left( A^T \lambda_A + C^T \lambda_C \right). 
\end{equation*}
Plugging the solution back into the constraint we get
\begin{equation*}
0 = A^T \lambda_A - P^T P \left( A^T \lambda_A + C^T \lambda_C \right) + C^T \lambda_C = \left( I - P^T P \right) \left( A^T \lambda_A + C^T \lambda_C \right). 
\end{equation*}
The reduced dual problem is given by
\begin{equation}
\label{eq:reduced_dual}
\begin{aligned}
\mathcal{D}_r: & \quad \min_{\lambda_A \in \mathbb{R}^I, \lambda_C \in \mathbb{R}^{q - \tilde{q}}} \lambda_A^Ts + f^{w*} \left( \begin{bmatrix} -P A^T & -P C^T \\ 0 & I \end{bmatrix} \begin{bmatrix} \lambda_A \\ \lambda_C \end{bmatrix} \right) \\
& \text{s.t.} \quad \left( I - P^T P \right) \left( A^T \lambda_A + C^T \lambda_C \right) = 0.
\end{aligned}
\end{equation}
The new dual affine constraint is exactly informed by the unobserved coordinates, since $I - P^T P$ projects onto these. The problem thus has the same dimension as the initial problem~\eqref{eq:general_raking}, and as many constraints as missing observations. We can easily apply Newton's method, particularly since it guarantees affine constraints are satisfied at each iteration. 

To recover $\beta^*$, we first recall that from the optimality conditions,
\begin{equation*}
\begin{aligned}
\zeta^* & = \nabla_z f^{w*} \left( \lambda_B^* \right) = \nabla_z f^{w*}\left( \begin{matrix} - P \left( A^T \lambda_A^* + C^T \lambda_C^* \right) \\ \lambda_C^* \end{matrix} \right) \\
P \beta^* & = \nabla_z f^* \left( -P \left( A^T \lambda_A + C^T \lambda_C \right) \right)
\end{aligned}
\end{equation*}
which means we can read off the solutions corresponding to those entries that were observed directly from evaluating the first component of the gradient of the dual objective in~\eqref{eq:reduced_dual}. This means we can rake observed values regardless of the missingness pattern. 

All that remains are the unobserved entries of $\beta$, which are precisely the Lagrange multipliers for the affine constraint of the reduced dual~\eqref{eq:reduced_dual}. Optimality of the associated Lagrangian with respect to $\lambda$ yields
\begin{equation*}
\begin{bmatrix} A \\ C \end{bmatrix} \left( I - P^T P \right) \beta^* = - \begin{bmatrix} s \\ 0 \end{bmatrix} - \begin{bmatrix} -A P^T & 0 \\ - C P^T & I \end{bmatrix} \nabla_z f^* \left( \begin{bmatrix} - P A^T & - P C^T \\ 0 & I \end{bmatrix} \begin{bmatrix} \lambda_A^* \\ \lambda_C^* \end{bmatrix} \right).
\end{equation*}
Thus we can rake missing observations exactly when the columns of the stack of $A$ and $C$ corresponding to unobserved coordinates of $\beta$ has full rank. 

We show in Section 3 of Supplement 3 how the raking of available observations and known aggregates allows imputing missing values while respecting the constraints.

\section{Application to Mortality Estimates}
\label{sec:application}

\begin{figure}
\begin{center}
\includegraphics[width=0.5\textwidth]{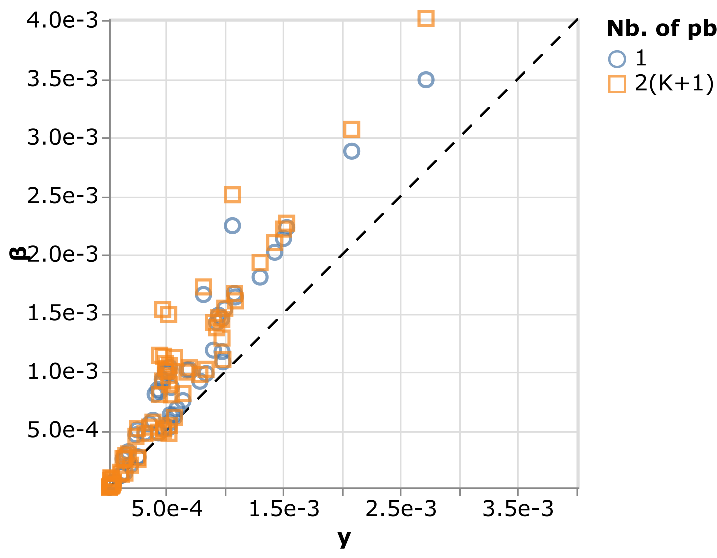}
\end{center}
\caption{Raked values vs initial rate of death for both methods for Delaware, age group 25-30. The raked values obtained with the approach proposed in this paper with a single minimization problem are usually closer to the initial observations than the raked values obtained with the previous approach with $2 \left( K + 1 \right)$ minimization problems.}
\label{fig:comparison_workflows}
\end{figure}

We first compare the results obtained with the previous method from~\cite{DWY_2016} and~\cite{DWY_2023} and obtained with the proposed method. In particular, we want to verify whether the raked values obtained when solving a single minimization problem are closer to the initial observations than the raked values obtained when solving the $2 \left( K + 1 \right)$ minimization problems. In Figure~\ref{fig:comparison_workflows}, we apply the raking procedure for Delaware, both-sexes, age group 25-to-30-year-old, using the entropic distance for $f$ and weights $w = 1$. We compare the reported values of the death rates (number of deaths divided by the population for each racial-ethnic group and county) with the corresponding raked values for both methods for all 72 cases (3 counties, 5 racial-ethnic groups and all racial-ethnic groups, 3 causes and all causes). The approach proposed in this paper with a single minimization problem gives raked values slightly closer to the initial values than the previous approach with $2 \left( K + 1 \right)$ minimization problems, meaning that the raked estimates stay closer to the original reported values while satisfying the requirements. This improvement is easier to see for larger reported death rates on the linear scale plot in Figure~\ref{fig:comparison_workflows}. 

\begin{figure}
\begin{center}
\includegraphics[width=\textwidth]{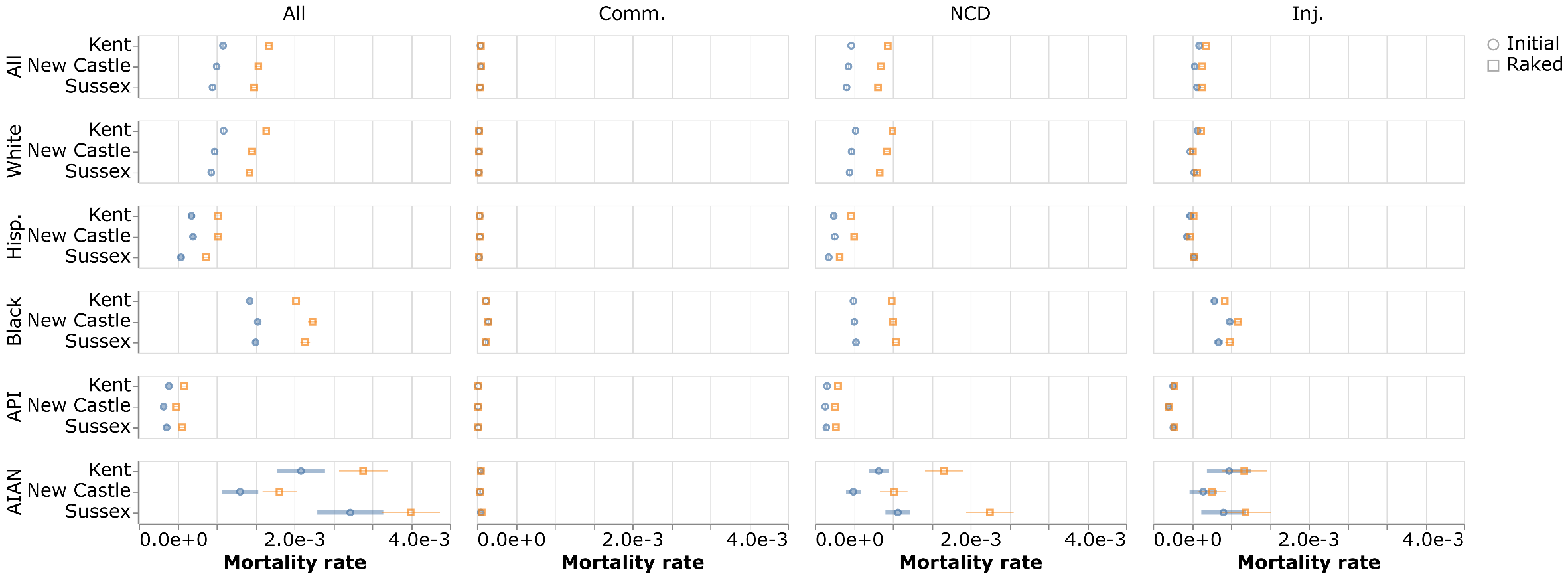}
\end{center}
\caption{Initial and raked values with associated uncertainty ordered by county. The raked values are close to the initial observations, but have smaller uncertainties as the uncertainty on the margins is smaller than the uncertainty reported for the observations.}
\label{fig:raked_county}
\end{figure}

\begin{figure}
\begin{center}
\includegraphics[width=\textwidth]{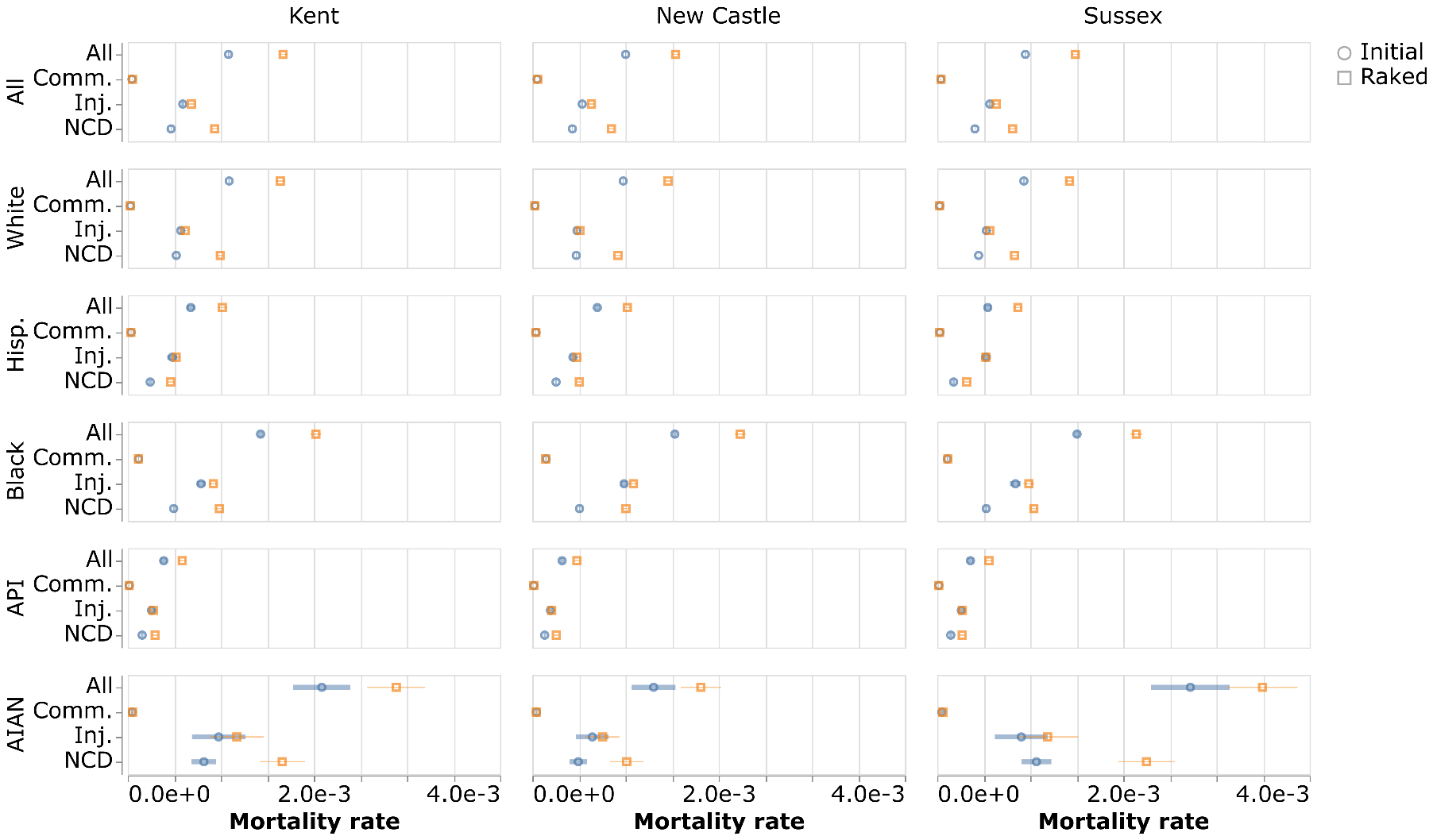}
\end{center}
\caption{Initial and raked values with associated uncertainty ordered by cause. The raked values are close to the initial observations, but have smaller uncertainties, as the uncertainty on the margins is smaller than the uncertainty reported for the observations.}
\label{fig:raked_cause}
\end{figure}

\begin{figure}
\begin{center}
\includegraphics[width=\textwidth]{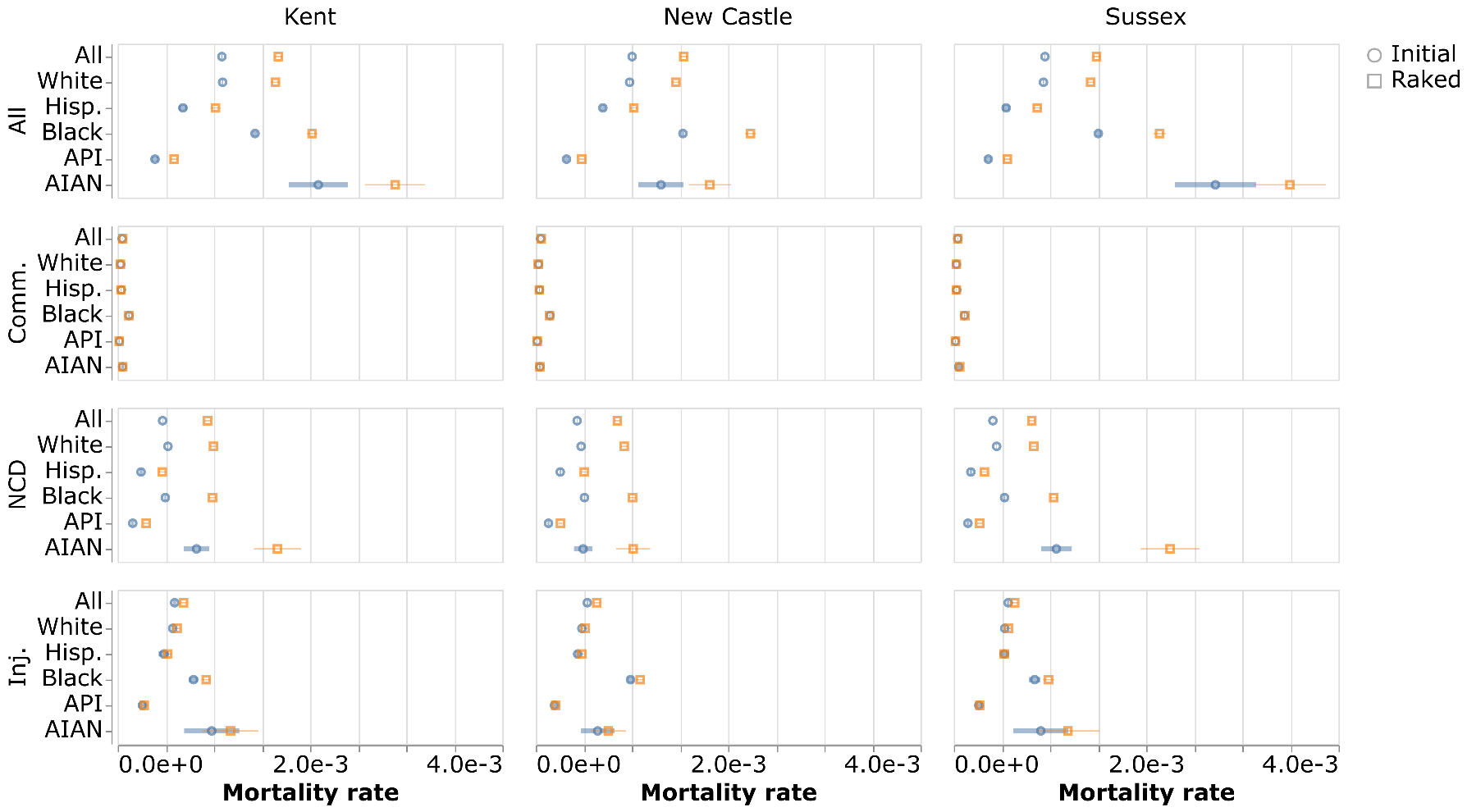}
\end{center}
\caption{Initial and raked values with associated uncertainty ordered by racial-ethnic group. The raked values are close to the initial observations, but have smaller uncertainties as the uncertainty on the margins is smaller than the uncertainty reported for the observations.}
\label{fig:raked_race}
\end{figure}

We compare the raked values with their associated uncertainty to the initial values. To better compare the different causes, racial-ethnic groups and counties on the same scale, we compare mortality rates. In Figure~\ref{fig:raked_county}, we look at the initial values and the raked values for both-sexes, Delaware, age group 25-to-30-year-old. The uncertainty is represented by a segment of length two standard deviations and centered on the estimated value. Races AIAN and API with much smaller population numbers have the largest uncertainties for both the initial and the raked values. This uncertainty on the raked values is significantly reduced compared to the uncertainty on the initial values. The initial uncertainty on the margins $s$ (that is on the GBD values) are smaller than the uncertainties on the observations $y$, and as a result the raking process significantly decreases the final uncertainty on these raked values.
We then show in Figures~\ref{fig:raked_cause} and~\ref{fig:raked_race} the initial and raked values with associated uncertainty ordered by cause and racial-ethnic group.

\begin{figure}
\begin{center}
\includegraphics[width=12cm]{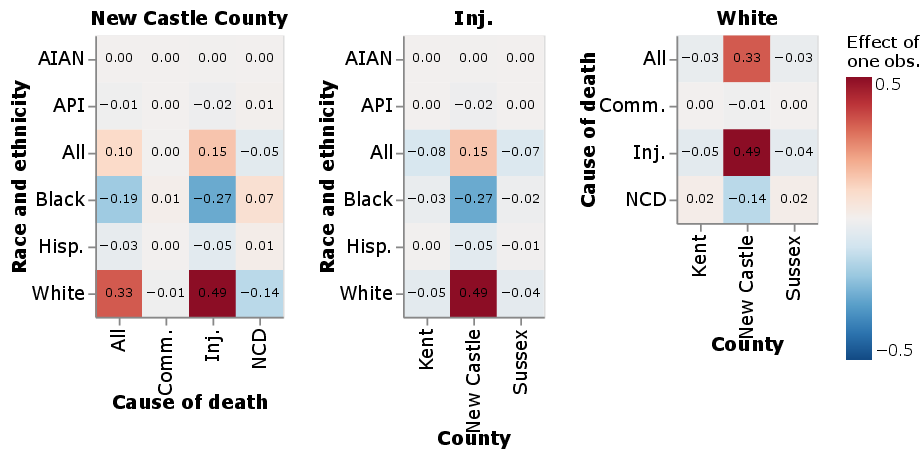}
\end{center}
\caption{Influence of the initial value of deaths for cause injuries, racial-ethnic group White and New Castle County on all the raked values. As expected, the corresponding raked value $\beta^*_{2,1,2}$ is the most affected, followed by the raked values for the partial totals $\beta^*_{0,1,2}$ and the $\beta^*_{2,0,2}$.}
\label{fig:most_initial}
\end{figure}

To estimate how the initial values influence the raked values, we look at the initial value for cause ``injuries'' (Inj), racial-ethnic group White and New Castle County (denoted $y_{2,1,2}$). Then we compute the corresponding values of the gradient $\frac{\partial \beta_{i,j,k}}{\partial y_{2,1,2}}$ for $i = 0 , \cdots , 3$, $j = 0 , \cdots , 5$ and $k = 1 , 2 , 3$. The corresponding values of the gradient are shown in Figure~\ref{fig:most_initial}. The most affected raked value is the one with the same cause, racial-ethnic group and county as the initial value considered. All causes deaths for racial-ethnic group White and New Castle County and all racial-ethnic groups deaths for cause ``injuries'' and New Castle County will be positively influenced by variations in the initial value $y_{2,1,2}$ while other causes and racial-ethnic groups will be negatively influenced. The raked values in the other counties will not be much affected by changes in the value of $y_{2,1,2}$.

\begin{figure}[h!]
\begin{center}
\includegraphics[width=\textwidth]{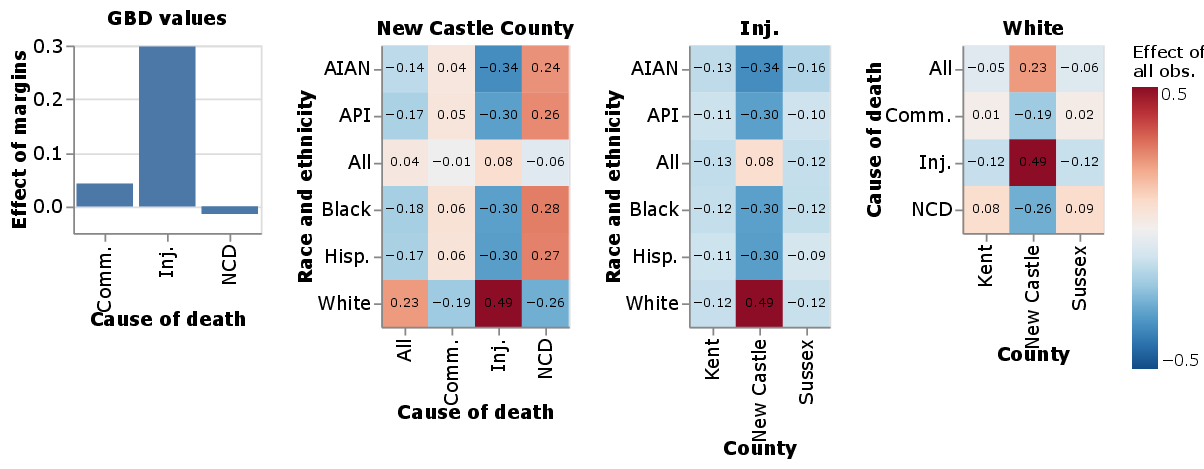}
\end{center}
\caption{Influence of all the initial values and aggregated initial values on the raked value for cause injuries, racial-ethnic group White and New Castle County. As expected, the corresponding observation $\bar{y}_{2,1,2}$ is the most influential, followed by the margin for the cause ``injuries''.}
\label{fig:most_raked}
\end{figure}

Using the intermediate derivative results, we can visualize which observation points have the most influence on the final estimated raked values. We look at the raked value for cause ``injuries'', racial-ethnic group White and New Castle County (denoted $\beta_{2,1,2}$). Then we compute the corresponding values of the gradient $\frac{\partial \beta_{2,1,2}}{\partial y_{i,j,k}}$ and $\frac{\partial \beta_{2,1,2}}{\partial s_i}$ for $i = 0 , \cdots , 3$, $j = 0 , \cdots , 5$ and $k = 1 , 2 , 3$. The corresponding values of the gradient are shown in Figure~\ref{fig:most_raked}. The most important observation point is the one with the same cause, racial-ethnic group and county as the raked value considered. Increases in all causes and all racial-ethnic groups deaths will increase the raked value, while increases in other causes and racial-ethnic groups will decrease it. The initial values for the other counties will have a smaller influence on the raked value. For the margins, increase in all racial-ethnic groups state mortality due to injuries will increase the raked value. An increase in all racial-ethnic groups state mortality due to communicable diseases  decreases the raked value, while an increase in all racial-ethnic groups state mortality due to non-communicable diseases increases the raked value as a result of increasing all cause deaths. 

\begin{figure}
\begin{center}
\includegraphics[width=6cm]{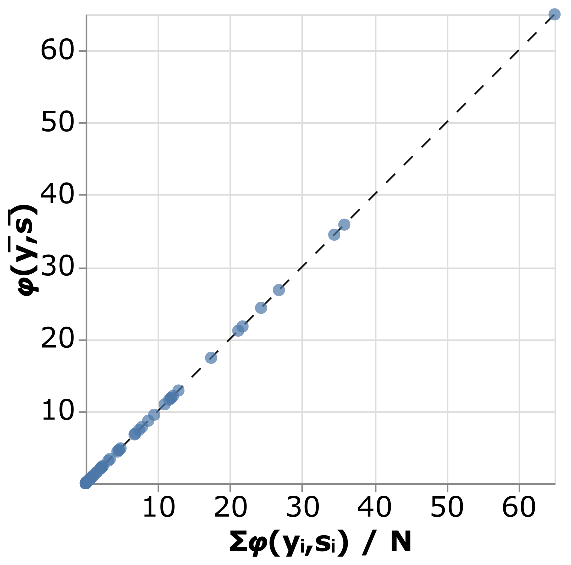}
\includegraphics[width=6cm]{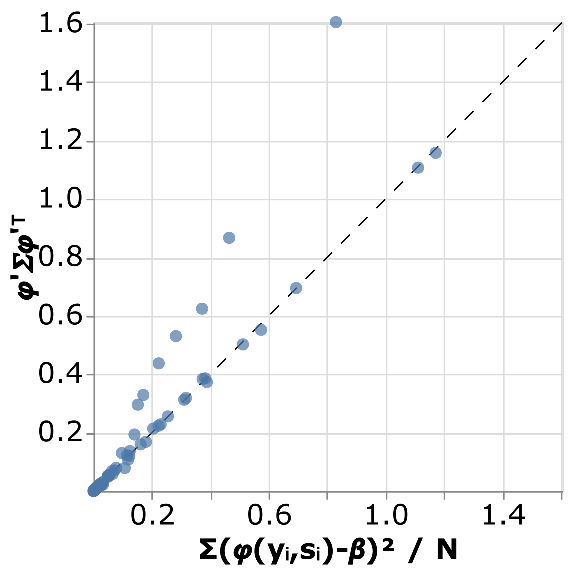}
\end{center}
\caption{Left: Comparison between the raked value corresponding to the sample mean of observations and margins and the sample mean of the raked values with 100 draws. Right: Comparison between the variance of the raked values estimated using the proposed uncertainty propagation method and the sample variance of the raked values with 100 draws. The new uncertainty propagation method gives more accurate results than using limited draws.} 
\label{fig:variance}
\end{figure}

Finally, we compute the variance of the estimated raked values using two methods: First, we take $100$ samples of the detailed and aggregate observations, apply the raking procedure to each sample, and compute the sample mean and the sample variance of the corresponding raked values. Second, we compute the sample mean of the initial data set, apply the raking procedure to the sample mean, and use the delta method and the implicit function theorem to propagate the variance of the observations to get the variance of the raked values. In Figure~\ref{fig:variance}, we look at the raked values for Delaware, both-sexes, and age group 25-to-30-year-old and compare the raked values and their standard deviation computed with both methods. We see evidence of sampling variation in the draws method, which makes sense since we only have 100 draws. The results obtained with the delta method are therefore both more accurate and far more efficient to compute compared to a Monte Carlo technique that rakes all draws.

\section{Conclusion}
\label{sec:conclusion}

In this paper, we have reviewed the convex optimization foundation of raking, and focused on a dual perspective that simplifies and streamlines prior raking extensions and provides new functionality, enabling a unified approach to $n$-dimensional raking, raking with differential weights, ensuring bounds on estimates, raking to margins either as hard constraints or as aggregate observations, handling missing data, and allowing efficient uncertainty propagation. The dual perspective also enables a fast and scalable matrix-free optimization approach for all of these extensions. We illustrated the capabilities using synthetic data and real mortality estimates. 


\begin{acks}[Acknowledgments]
An accompanying Python package can be found and installed through PyPI (\url{https://pypi.org/project/raking/}). It is also available on GitHub (\url{https://github.com/ihmeuw-msca/raking}). The Python scripts used to make the figures can be found on the Github account of the first author (\url{https://github.com/ADucellierIHME/Raking_paper}).
\end{acks}

\begin{funding}
This work was funded by the Bill and Melinda Gates Foundation.
\end{funding}

\begin{supplement}
\stitle{1. Historical context}
\sdescription{In this supplement, we describe the historical context of raking in the survey sampling domain and detail the closed form solution of the 1D problem and the IPF algorithm.}
\end{supplement}
\begin{supplement}
\stitle{2. Technical details and proofs}
\sdescription{In this supplement, we give additional technical details and proofs for the methods section.}
\end{supplement}
\begin{supplement}
\stitle{3. Additional verifications}
\sdescription{In this supplement, we illustrate with examples the raking methods described in Section~\ref{sec:discussion}.}
\end{supplement}
\begin{supplement}
\stitle{4. User interface}
\sdescription{In this supplement, we describe the user interface for the accompanying Python package.}
\end{supplement}


\bibliographystyle{imsart-nameyear} 
\bibliography{main_paper} 

\begin{frontmatter}
\title{Historical context}
\runtitle{Historical context}
\end{frontmatter}

\setcounter{section}{0}
\section{Historical context}

We review the historical context of the raking problem in survey sciences and global health. Raking originated in survey inference, where it was used to adjust the sampling weights of the cases in a contingency table in order to match the marginal totals of the adjusted weights to the corresponding totals for the population (cf. Till\'e, 2020). Given a population $\mathcal{U}$ we want the population total of a variable $Y$, 
\(
\displaystyle 
\mathcal{Y} = \sum_{i \in \mathcal{U}} y_i, 
\)
estimated using a sample $S \subseteq \mathcal{U}$. Let $\pi_i$ 
denote the inclusion probability of individual $i$ into the sample,
and let $d_i = \frac{1}{\pi_i}$. The classic Horvitz-Thompson-Narain estimator
for $\mathcal{Y}$ (Narain, 1951; Horvitz and Thompson, 1952) is given by 
\begin{equation}
\label{eq:estimator}
\hat{\mathcal{Y}} = \sum_{i \in S} \frac{y_i}{\pi_i} = \sum_{i \in S} d_i y_i.
\end{equation}
We may also have access to information on $k$ auxiliary variables. We define by $x_i \in \mathbb{R}^k$ the vector of the values taken by the $k$ auxiliary variables for an individual $i$. While individual auxiliary vectors are unknown, suppose we have the marginal sum $s \in \mathbb{R}^k$ from the population $\mathcal{U}$. The goal of raking in survey science is to replace the weights $d_i$ in~\eqref{eq:estimator} by raking weights $w_i$ such that
\begin{equation*}
\sum_{i \in S} w_i x_i = s,
\end{equation*}
and use these to compute an estimate of $\mathcal{Y} = \sum_{i\in S} w_i y_i$ with associated uncertainty. 

The formulations and goals of raking in global health research are different. We consider the sample $S$ to be the population $\mathcal{U}$, and adjust the values taken by the $y_i$ so their sum is equal to the known total $s$, which may come from a trusted data source or model of lower granularity. To map this case to the above framework,  the weights $d_i$ are identically equal to $1$, while auxiliary variables $x$ are different margins over $\mathcal{U}$. For example, in one-dimensional raking,  there is a single auxiliary variable and corresponding marginal total $\sum_{i \in \mathcal{U}} x_i = s$. In 2-dimensional raking, auxiliary variables are row and column sums of $Y$, with corresponding marginal vector $s$. In all cases, rather than looking for raking weights $w_i$, we are interested in directly finding the raked values $\beta_i = w_i y_i$.

\section{1-D raking}

We consider the classic one-dimensional case that motivates the use of entropic distance. The raking problem can be framed as minimization of a separable function subject to a simple equality constraint, with the corresponding dual:
\begin{equation*}
\mathcal{P}: \quad \min_{\beta \in \mathbb{R}^p} \sum_{i = 1}^p f_i \left( \beta_i ; y_i \right) \quad \text{s.t.} \quad \mathbb{1}^T \beta = s \quad \Longleftrightarrow \quad \min_{\lambda \in \mathbb{R}} \lambda s + \sum_{i = 1}^p f_i^* \left( - \lambda \right) \quad :\mathcal{D}
\end{equation*}
with optimal solution pair $\left( \beta^* , \lambda^* \right)$:
\begin{equation*}
\beta^* = \nabla_z f^* \left( - \lambda \mathbb{1}_p \right) \quad \text{or} \quad \beta^*_i = \nabla_z f^*_i \left( - \lambda \right).
\end{equation*}
For the entropic distance function, we have
\begin{equation*}
f_i \left( \beta_i , y_i \right) = \beta_i \ln \left( \frac{\beta_i}{y_i} \right) - \left( \beta_i - y_i \right) \quad \Longleftrightarrow \quad f_i^* \left( z \right) = y_i \left( \exp \left( z \right) - 1 \right),
\end{equation*}
and so the optimal solution pairs simplifies to
\begin{equation*}
\beta^* = y \odot \exp \left( - \lambda \mathbb{1}_p \right) \quad \text{or} \quad \beta^*_i = y_i \exp \left( - \lambda \right)
\end{equation*}
By taking the derivative of the dual problem, we finally get
\begin{equation*}
s = \sum_{i = 1}^p y_i \exp \left( - \lambda \right) \quad \Rightarrow \quad \exp \left( - \lambda \right) = \frac{s}{\sum_{i= 1}^p  y_i}, \quad \beta^* = y \odot \exp \left( - \lambda \mathbb{1}_p \right) = \frac{s}{\sum_{i = 1}^p y_i} y.
\end{equation*}
This computation recovers the classic bedrock result that 1D raking with entropic distance scales observations by the ratio of the constraint and sum.

\section{Iterative Proportional Fitting (IPF) algorithm}

A key connection to optimization appears in two-dimensional raking, when we are given  linear combinations of the data (e.g. $n$ row sums and $m$ column sums) as constraints. We then have $k = m + n - 1$ constraints, as the $\left( m + n \right)$th constraint is a linear combination of the previous ones. If in problem (7) from the main text, we choose $B = I_p$ and $w = \mathbb{1}_p$, the raking problem takes the form
\begin{equation}
\label{eq:2DRake}
\min_\mathbf\beta f \left( \mathbf\beta ; Y \right) \quad \text{s.t.} \quad \mathbf\beta \mathbb{1}_n = s_r \text{ and } \mathbf\beta^T \mathbb{1}_m = s_c,
\end{equation}
where we use a matrix representation of $\mathbf\beta, Y \in \mathbb{R}^{m \times n}$ instead of vectors from $\mathbb{R}^{m n}$ as in problem (7), $s_r \in \mathbb{R}^m$ and $s_c \in \mathbb{R}^n$. The iterative proportional fitting (IPF) algorithm, an algorithm to solve \eqref{eq:2DRake} with $f$ the entropic distance function
\begin{equation*}
f \left( \beta; y \right) = \sum_{i=1}^m \sum_{j=1}^n \beta_{i,j} \log\left(\frac{\beta_{i,j}}{y_{i,j}}\right) - \left(\beta_{i,j} - y_{i,j} \right)
\end{equation*}
was first proposed by Deming and Stephan (1940) and further developed by Stephan (1942), though these early works focused on the iterative raking procedure rather than the underlying formulation~\eqref{eq:2DRake}. The calibration method was further developed by Deville (1992) and Deville \textit{et al.} (1993) to adjust samples on known population totals. A review of the method and optimization formulations is found in Devaud and Till\'e (2019), with the exception of the explicit dual and primal-dual relationships developed here.  

A great reference to the IPF method itself from an optimization perspective is provided by She and Tang (2019). IPF and the equivalent, the Sinkhorn iteration (Sinkhorn, 1967) take advantage of block-separable constraints for rows and columns to obtain efficient primal-dual updates in 2D raking and related problems, such as optimal transport (Cuturi, 2013). In the current paper, we show that the same underlying structure can be used in simple direct methods, such as Newton, to solve the dual just as efficiently. This allows us to focus on generalizing the model to apply to extended use cases, such as inexact observations rather than marginal constraints (Williams and Savitsky, 2024), to forego the development of modified IPF algorithms, and to rely on the sparsity of the underlying linear operators to maintain computational efficiency.

Contrary to the one-dimensional case below, the two-dimensional entropic raking does not admit a closed form solution. Instead, we use the special structure of the 2D raking problem to understand the IPF algorithm from a dual perspective. 

The 2D raking problem~\eqref{eq:2DRake} with unknown $\mathbf{\beta} \in \mathbb{R}^{m \times n}$ can be written as: 
\begin{equation}
\label{2d_raking_full}
\begin{aligned}
\mathcal{P}: & \quad \min_{\mathbf{\beta}} f \left( \mathbf{\beta} ; Y \right) \quad \mbox{s.t} \quad \mathbf{\beta} \mathbb{1}_n = s_r, \quad \mathbf{\beta}^T \mathbb{1}_m = s_c \\
\mathcal{L}: & \quad f \left( \mathbf{\beta} \right) + \lambda_r^T \left( \mathbf{\beta} \mathbb{1}_n - s_r \right) +  \lambda_c^T \left( \mathbf{\beta}^T \mathbb{1}_m -s_c \right) \\
\mathcal{D}: & \quad 
\min_{\lambda_r, \lambda_c} \lambda_c^T s_c + \lambda_r^T s_r + \sum_{i,j} f_{ij}^* \left( -\lambda_r^i - \lambda_c^j \right). \\
\text{OPT}: & \quad \mathbf\beta^* = \nabla_z f^*\left( - \lambda_r^* \otimes \lambda_c^{*T} \right) \quad \text{or} \quad \beta^*_{ij} = \nabla_z f^* \left( - \lambda_r^{*i} - \lambda_c^{*j} \right).
\end{aligned}
\end{equation}
The 2D structure is explicitly used to frame $\lambda_r \in \mathbb{R}^m$ and $\lambda_c \in \mathbb{R}^n$ as `row' and `column' multiplier vectors. 

The IPF iteration and equivalent Sinkhorn algorithm take advantage of the fact that for the entropic distance, we have the conjugate
\begin{equation*}
\nabla_z f_{ij}^* \left( - \lambda_r^i - \lambda_c^j \right) = y_{ij} \exp \left( - \lambda_r^i \right) \exp \left( - \lambda_c^j \right)
\end{equation*}
with two blocks $\lambda_r$ and $\lambda_c$ and dual objective that is separable with respect to one block when the other is fixed. Moreover, partial minimizers for each block are available in closed form, allowing a block-coordinate update:
\begin{eqnarray*}
\exp \left( - \lambda_r^i \right)^+ & = & s_r^i / \sum_j y_{ij} \exp \left( -\lambda_c^j \right) \quad \text{for} \quad i = 1, \dots, m \\
\exp \left( -\lambda_c^j \right)^+ & = & s_c^j / \sum_i y_{ij} \exp \left( -\lambda_r^i \right)^+ \quad \text{for} \quad j = 1, \dots, n. 
\end{eqnarray*}
While the IPF is typically viewed as iterative scaling for rows and columns of $\mathbf{\beta}$ to satisfy alternating sets of primal constraints, in light of~\eqref{2d_raking_full} it is also an alternating maximization method for the dual. The updates are fully independent within each block, so they can be parallelized across the $m$ rows and $n$ columns. These key findings, together with the impressive performance of block coordinate descent, explain the emphasis of previous literature on the IPF and Sinkhorn (Deming and Stephan (1940), Stephan (1942), She and Tang(2019), Cuturi (2013)).

For the two-dimensional raking case, we can make the Hessian-vector product structure more apparent. We first rewrite the row and column constraints as a single set of equality constraints $A\beta = s$. Define $\beta = \text{vec} \left( \mathbf\beta \right)$ and recall the classic vec-kron identity
\begin{equation}
\label{eq:vec-kron}
\text{vec} \left( ABC \right) = \left[ C^T \otimes A \right] \text{vec} \left( B \right).
\end{equation}
Using~\eqref{eq:vec-kron} we can directly write
\begin{equation}
\label{eq:explicit2D}
\begin{aligned}
\text{vec} \left( I_m \mathbf{\beta}\mathbb{1}_n \right) = \left( \mathbb{1}_n^T \otimes I_m \right) \beta & \qquad
A = \begin{bmatrix}
\mathbb{1}_n^T \otimes I_m \\
I_n \otimes \mathbb{1}_m^T
\end{bmatrix} \\
\text{vec} \left( \mathbb{1}_m^T \mathbf{\beta} I_n \right) = \left( I_n \otimes \mathbb{1}_m^T \right)\beta & \qquad s = \begin{bmatrix} s_r \\ s_c\end{bmatrix}.
\end{aligned}
\end{equation}
and the dual in~\eqref{2d_raking_full} can be written as
\begin{equation*}
\mathcal{D}: \quad \min_{\lambda_c, \lambda_r} \lambda_c^T s_c + \lambda_r^T s_r + f^* \left( - \left( \mathbb{1}_n \otimes I_m \right) \lambda_r - \left ( I_n \otimes \mathbb{1}_m \right) \lambda_c \right). 
\end{equation*}
The Hessian of the dual objective is given by
\begin{equation*}
H = A S A^T, \quad S = \nabla_z^2 f^* \left( - \left( \mathbb{1}_n \otimes I_m \right) \lambda_r - \left( I_n \otimes \mathbb{1}_m \right) \lambda_c \right)
\end{equation*}
where $S$ is diagonal since $f$ is separable across coordinates, and $A \in \mathbb{R}^{\left( n + m \right) \times \left(n \cdot m \right)}$. The separability between row and column blocks is manifest in the structure of $H$:
\begin{equation*}
A S A^T = 
\begin{bmatrix}
\mathbb{1}_n^T \otimes I_m \\
I_n \otimes \mathbb{1}_m^T
\end{bmatrix}
S
\begin{bmatrix}
\mathbb{1}_n \otimes I_m &
I_n \otimes \mathbb{1}_m
\end{bmatrix}
=
\begin{pmatrix} \left[ \mathbb{1}_n^T \otimes I_m \right] S \left[ \mathbb{1}_n \otimes I_m \right] & \left[ \mathbb{1}_n^T \otimes I_m \right] S \left[ I_n \otimes \mathbb{1}_m \right] \\ \left[ I_n \otimes \mathbb{1}_m^T \right] S \left[ \mathbb{1}_n \otimes I_m \right] & \left[ I_n \otimes \mathbb{1}_m^T \right] S \left[ I_n \otimes \mathbb{1}_m \right] \end{pmatrix}.
\end{equation*}
The top left and bottom right blocks are diagonal:
\begin{equation*}
\left( \mathbb{1}_n^T \otimes I_m \right) S \left( \mathbb{1}_n \otimes I_m \right) = \sum_{i = 1}^n S_i, \quad  \left[ I_n \otimes \mathbb{1}_m^T \right] S \left[ I_n \otimes \mathbb{1}_m \right] =  \sum_{i = 1}^m \left( P S \right)_i
\end{equation*}
where $S_i\in\mathbb{R}^{m\times m}$ are diagonal blocks of $S\in\mathbb{R}^{mn \times  mn}$, and $P$ is a permutation matrix that rearranges the order of the $n + m$ diagonal entries to follow row rather than column order, e.g. $P \left( \text{diag} \left( s_1 , s_2 , s_3 , s_4 , s_5 , s_6 \right) \right) = \text{diag} \left( s_1,  s_4 , s_2 , s_5 , s_3 , s_6 \right)$ for $m = 3, n = 2$. The IPF algorithm takes advantage of this explicit structure to implement block coordinate descent.

The key fact is that Hessian-vector products have $O \left( n m \right)$ complexity, just like IPF iterations:
\begin{equation}
\label{eq:explicitAction2D}
A S A^T x = A \underbrace{\text{diag} \left( S \right) \odot \left(\text{vec} \left( x_{1:n} \mathbb{1}_m^T \right) + \text{vec} \left( \mathbb{1}_n x_{n + 1 : n + m}^T \right) \right)}_z = \begin{bmatrix} Z \mathbb{1}_n \\ Z^T \mathbb{1}_m \end{bmatrix}
\end{equation}
where $Z\in\mathbb{R}^{m\times n}$ is the vec adjoint (i.e. reshape) of $z \in \mathbb{R}^{n m \times 1}$. 

\begin{figure}[h!]
\includegraphics[width=0.65\linewidth]{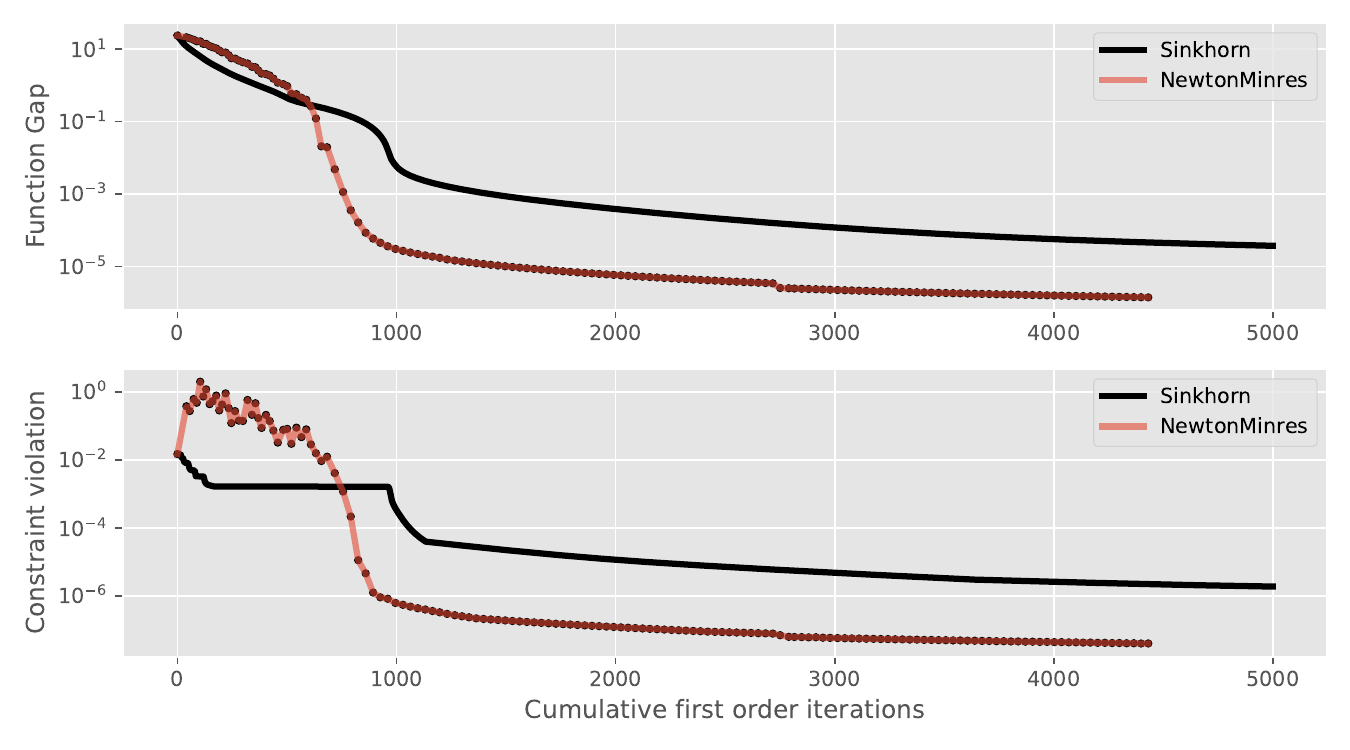}
\caption{Sinkhorn/IPF iteration (black) vs. inexact Newton on the dual in terms of cumulative matrix vector products (x-axis). Top panel shows function gap on the dual objective, while bottom panel shows the maximum constraint violation. The Newton algorithm uses the iterative Minres subsolver, relying only on efficient Hessian-vector products, and solves to a relative tolerance of $10^{-4}$ with a maximum of 100 iterations. We set $n=300$, $m=200$, $s_r = \frac{1}{n}\mathbb{1}_n, s_c = \frac{1}{m}\mathbb{1}_m$, and $\log(y_{ij}) \stackrel{iid}\sim N(0,200^2)$.}
\label{fig:ntmr}
\end{figure}

While~\eqref{eq:explicitAction2D} gives insight into the 2D case, we do not need the specificity of ~\eqref{eq:explicit2D} to get optimal $O \left( n m \right)$ complexity. All we need to know is that the number of nonzero entries in $A$ for the 2D case is $O \left( n m \right)$, since in the classic case we have row and column constraints. Then it is immediate that applying $A^T$, scaling by $S$, and applying $A$ are each $O \left( n m \right)$ operations. We compare the specialized IPF method for the 2D case to an inexact Newton method using the Minres subsolver in Figure~\ref{fig:ntmr}. Inexact Newton  works on the dual, and relies on problem structure only sparsity of $A$; it is is agnostic to the full specificity of~\eqref{eq:explicit2D}. Nonetheless, it outperforms IPF both in dual gap and constraint violation.

As we generalize the raking framework, we favor the inexact Newton approach that seamlessly leverages sparsity of linear operators, and avoids the need for specialized IPF development for each innovation.

\begin{frontmatter}
\title{Technical details and proofs}
\runtitle{Technical details and proofs}
\end{frontmatter}

\setcounter{section}{0}
\section{Convex preliminaries}
\label{sec:convex_pre}

We present selected topics from convex analysis that we need to frame the dual perspective on raking. For a canonical reference that goes much deeper into these ideas, please see Rockafellar and Wets (2009).

For a function $f:\mathbb{R}^p\rightarrow \overline{\mathbb{R}}:= \mathbb{R}\cup \{-\infty, \infty\}$, define the {\it epigraph}
\begin{equation*}
\text{epi} \left( f \right) = \{ \left( \beta ,\alpha \right) : f \left( \beta \right) \leq \alpha\} \subset \mathbb{R}^{p+1}. 
\end{equation*}
A function $f : \mathbb{R}^p\rightarrow \overline{\mathbb{R}}$ is {\it closed} when $\text{epi} \left( f \right)$ is a closed set, {\it proper} when $f\not\equiv\infty$  and does not take the value of $-\infty$, and {\it convex} when $\text{epi} \left( f \right)$ is a convex set.

\subsection{Convex conjugate}

For any function $f : \mathbb{R}^p\rightarrow \overline{\mathbb{R}}$, its {\it convex conjugate} is defined by
\begin{equation*}
f^* \left( z \right) = \sup_\beta z^T \beta - f \left( \beta \right).
\end{equation*}
From the definition, {\it Fenchel's inequality} is immediate:
\begin{equation}
\label{eq:fenchelIneq}
f \left( \beta \right) + f^* \left( z \right) \leq z^T \beta \quad \forall \beta , z.
\end{equation}
For closed convex $f$, we have $f^{**} = f$, and we can write
\begin{equation}
\label{eq:symmFenchel}
f \left( \beta \right) = f^{**} \left( \beta \right) = \sup_z z^T\beta - f^* \left( z \right). 
\end{equation}
When $f$ and $f^*$ are differentiable, and we have attainment in the defining equation~\eqref{eq:fenchelIneq}, we can differentiate each equality to get a key relationship between $\nabla f$ and $\nabla f^*$:
\begin{equation}
\label{eq:fenchelEq}
z = \nabla_\beta f \left( \beta \right) , \quad \beta = \nabla_z f^* \left( z \right) \quad \Rightarrow \quad \nabla_z f^* \left( \cdot \right) = \nabla_\beta f \left( \cdot \right)^{-1}.
\end{equation}
For practical computation, the following basic identity relates conjugates of linear transforms of $f$ to linear transforms of $f^*$:
\begin{equation*}
\left( c f \left( \left( \beta - a \right) / b \right) \right)^* \left( z \right) = c f^* \left( b z / c \right) + a z. 
\end{equation*}
Finally, conjugates of separable functions are separable: 
\begin{equation*}
f \left( \beta \right) = \sum_i f_i \left( \beta_i \right) \quad \Rightarrow \quad f^* \left( z \right) = \sum_i f_i^* \left( z_i \right). 
\end{equation*}

\subsection{Convex duality}

The notion of {\it convex duality} is closely related to conjugacy. Given any convex program
\begin{equation}
\label{eq:primal}
\mathcal{P}: \quad \inf_\beta f \left( \beta \right) + g \left( A \beta - b \right) + c^T \beta
\end{equation}
where $f , g$ are closed proper convex functions, we can associate a saddle point system by writing $g$ in terms of its conjugate~\eqref{eq:symmFenchel}:
\begin{equation*}
g \left( A \beta - b \right) = g^{**} \left( A \beta - b \right) = \sup_\lambda \lambda^T \left( A \beta - b \right) - g^* \left( \lambda \right). 
\end{equation*}
Substituting into~\eqref{eq:primal}, we define the resulting object as the Lagrangian:
\begin{equation}
\label{eq:Lagrangian}
\mathcal{L} \left( \beta, \lambda \right) := f \left( \beta \right) + c^T \beta + \lambda^T \left( A \beta - b \right) - g^* \left( \lambda \right). 
\end{equation}
While the primal is obtained by taking $\sup_\lambda \mathcal{L} \left( \beta , \lambda \right)$, we can compute the dual by interchanging the order and taking the infimum first:
\begin{equation}
\label{eq:dual}
\begin{aligned}
\mathcal{D}: & \quad \sup_\lambda \inf_\beta f \left( \beta \right) + c^T \beta + \lambda^T \left( A \beta - b \right) - g^* \left( \lambda \right) \\
& = \sup_\lambda - \lambda^T b - g^* \left( \lambda \right)  - \sup_\beta \left( - c - A^T \lambda \right)^T \beta - f \left( \beta \right) \\
& = \sup_\lambda - \lambda^T b - g^* \left( \lambda \right) - f^* \left( - c - A^T \lambda \right).
\end{aligned}
\end{equation}
Interchanging the order of optimization means the primal value is greater than or equal to the dual value. Simple technical conditions guarantee equality of primal and dual values and allow for optimal solutions to be characterized by simple equations and inequalities (Rockafellar and Wets, 2009). We focus on the relevant case in the next section. 

\subsection{Differentiable objective with equality constraints}

For the special case when $f$ and $f^*$ are differentiable and $g$ encodes equality constraints, we can take
\begin{equation*}
g \left( u \right) = \delta_0 \left( u \right) = \begin{cases}
0 & u = 0 \\
\infty & u \neq 0.
\end{cases}
\end{equation*}
The conjugate of the indicator function to $0$ is then given by
\begin{equation*}
g^* \left( z \right) = \sup_u u^T z - g \left( u \right) = 0. 
\end{equation*}
That means the dual program~\eqref{eq:dual} simplifies to
\begin{equation*}
\sup_\lambda -\lambda^T b - f^* \left( - c - A^T \lambda \right) \quad \Leftrightarrow \quad \min_\lambda \lambda^T b + f^* \left( - c - A^T \lambda \right),
\end{equation*}
where we have written the dual as a minimization problem for convenience. The dual objective in the latter variant is smooth and convex, and the dual is unconstrained. Moreover, the Lagrangian~\eqref{eq:Lagrangian} simplifies to
\begin{equation*}
\mathcal{L} \left( \beta , \lambda \right) = f \left( \beta \right) + c^T \beta + \lambda^T \left( A \beta - b \right)
\end{equation*}
and we can obtain primal-dual relationships by differentiating the Lagrangian:
\begin{equation}
\label{eq:opt_cond}
\nabla_\beta f \left( \beta^* \right) + c + A^T \lambda^* = 0 \quad  \Rightarrow \quad \beta^* = \left( \nabla_\beta f \right)^{-1} \left( - c -A^T \lambda^* \right) = \nabla_z f^* \left( - c - A^T \lambda^* \right),
\end{equation}
where we have used~\eqref{eq:fenchelEq} to pass from the inverse of the gradient map to the gradient of the conjugate. Thus once we solve the dual, we can recover the primal solution simply by evaluating the gradient of the conjugate, regardless of the form of $f$.  

\section{Proof of Theorem 2.1}
\label{sec:proof}

Here, we detail the proof of Theorem 2.1, given in Section 2.2.1. From the central limit theorem, we have
\begin{equation*}
\sqrt{n} \left( \begin{pmatrix} \bar{y}_n \\ \bar{s}_n \end{pmatrix} - \theta \right) \rightarrow \mathcal{N} \left( 0 , \Sigma \right), \quad 
\text{where} \quad 
\theta = \mathbb{E} \begin{pmatrix} y^{(i)} \\ s^{(i)} \end{pmatrix}, \quad 
\Sigma = \begin{pmatrix} \Sigma_y & \Sigma_{ys} \\ \Sigma_{ys} & \Sigma_s \end{pmatrix}.
\end{equation*}
If the  mapping $\phi: \mathbb{R}^{q + I}\rightarrow \mathbb{R}^{p + q}$ from detailed and aggregate observations to the raked estimates is differentiable at $\theta$, then we have that $\sqrt{n} \left( \phi \begin{pmatrix} \bar{y}_n \\ \bar{s}_n \end{pmatrix} - \phi \left( \theta \right) \right)$ converges weakly to a multivariate normal distribution with mean $0$ and covariance $\phi'_\theta \Sigma \phi^{'T}_\theta$.

\section{Conjugate of logistic loss}
\label{sec:conjugate_logit}

In this part, we detail the computation of the conjugate of the logistic loss given in Section 3.2.3. To compute the conjugate, we drop the index for convenience and define
\begin{equation*}
\tilde{\beta} = \frac{\beta - l}{u - l}, \quad \tilde{y} = \frac{y - l}{u - l}, \quad  f \left( \beta \mid y \right) = \left( u - l \right) \left( \tilde{\beta} \log \left( \frac{\tilde{\beta}}{\tilde{y}} \right) + \left( 1 - \tilde{\beta} \right) \log \left( \frac{1 - \tilde{\beta}}{1 - \tilde{y}} \right) \right).
\end{equation*}
Next, we compute the conjugate of
\begin{equation*}
\tilde{f} \left( x \right) = x \log \left( \frac{x}{y} \right) + \left( 1 - x \right) \log \left( \frac{1 - x}{1 - y} \right).
\end{equation*}
We have
\begin{equation*}
\tilde{f}^* \left( z \right) = \sup_x \left[ x z - x \log \left( \frac{x}{y} \right) - \left( 1 - x \right) \log\left( \frac{1 - x}{1 - y} \right) \right].
\end{equation*}
Taking the derivative with respect to $x$, we get
\begin{equation*}
z = \log \left( \frac{x}{y} \right) - \log \left( \frac{1 - x}{1 - y} \right),
\end{equation*}
which gives us
\begin{eqnarray*}
\tilde{f}^* \left( z \right) & = & x \left( \log \left( \frac{x \left( z \right)}{y} \right) - \log \left( \frac{1 - x \left( z \right)}{1 - y} \right) \right) - x \left( z \right) \log \left( \frac{x \left( z \right)}{y} \right) - \left( 1 - x \left( z \right) \right) \log \left( \frac{1 - x \left( z \right)}{1 - y} \right) \\
& = & - \log \left( \frac{1 - x \left( z \right)}{1 - y} \right) = z - \log \left( \frac{x \left( z \right)}{y} \right), \quad x \left( z \right) =  \frac{y}{y + \left( 1 - y \right)\exp \left( - z \right)}.
\end{eqnarray*}
Thus, we have
\begin{equation*}
\tilde{f}^* \left( z \right) = z + \log \left( y + \left( 1 - y \right) \exp \left( - z \right) \right) = \log \left( y \exp \left( z \right) + \left( 1 - y \right) \right).
\end{equation*}
and the final conjugate is
\begin{eqnarray*}
f^* \left( z \right) & = & \left( u - l \right) \log \left( \tilde{y} \exp \left( z \right) + \left( 1 - \tilde{y} \right) \right) + l z \\
& = & \left( u - l \right) \log \left( \frac{y - l}{u - l} \exp \left( z \right) + \frac{u - y}{u - l} \right) + l z.
\end{eqnarray*}

\begin{frontmatter}
\title{Additional verifications}
\runtitle{Additional verifications}
\end{frontmatter}

In this section, we illustrate the raking extensions described in Section 3 on simple examples. 

\setcounter{section}{0}
\section{Impact of differential weights}
\label{sec:example_weights}

\begin{figure}[h!]
\begin{center}
\includegraphics[width=0.6\textwidth]{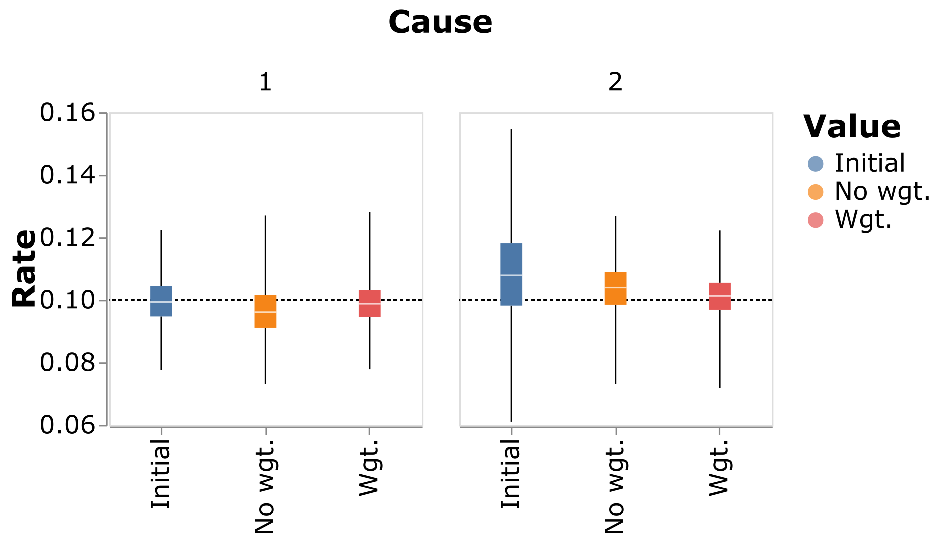}
\end{center}
\caption{Distribution of initial values (blue), raked values without weights (orange) and raked values with weights (red) for the unbiased cause $1$ and the biased cause $2$. Adding weights to the raking allows to get a final result closer to the known true value of the rate.}
\label{fig:simulation_weights}
\end{figure}

To illustrate the raking weights approach explained in Section 3.1, we rake death rates from two different causes to the all-cause death rate margin, known to be $0.2$. For each cause, death rate is computed from $10$ observations that follow a binomial distribution. For cause $1$, the probability of success $p_1$ of the binomial distribution is equal to $0.1$ and the number of samples varies uniformly between $100$ and $200$. For cause $2$, the observation mechanism is noisy and biased, with probability of success $p_2$ set  to $\text{expit} \left( \text{logit} \left( 0.1 \right) + \mathcal{N} \left( 0 , 0.5 \right) \right)$. Results for $500$ simulations are presented in Figure~\ref{fig:simulation_weights}. For each simulation, we rake the rate of death without weights and with weights equal to the inverse of the variance of the calculated rates of death.  The initial value for cause $1$ is centered around its true value $0.1$. However, the initial value for cause $2$ is overestimated. When raking without weights, both values are multiplied by the same ratio in order to match the margins value. The raked value for cause $1$, which was more certain, is thus farther away from its true value than before the raking process. When raking with weights, the more certain cause $1$ is assigned a bigger weight than the more uncertain biased cause $2$. After raking, cause $1$ stays close to its true value and cause $2$ gets closer to its true value than in the case when we rake without weights.

\section{Impact of raking loss}
\label{sec:example_loss}

\begin{figure}[h!]
\begin{center}
\includegraphics[width=0.6\textwidth]{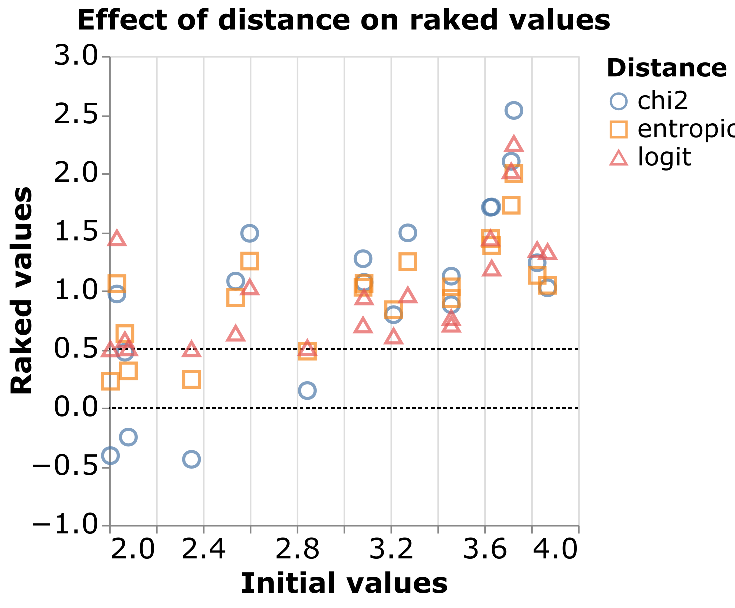}
\end{center}
\caption{Raked values versus initial values when raking with the $\chi$-square loss (blue circles), the entropic loss (orange squares) and the logistic loss (red triangles). The entropic loss ensures that all raked values remain positive. The logistic loss ensures that all raked values remain bounded in the interval $\left[ 0.5 ; 4 \right]$.}
\label{fig:distance_effect}
\end{figure}

To illustrate the effect of the choice of raking loss discussed in Section 3.2 on the final raking values, we generated a $4 \times 5$ 2D synthetic dataset with observations following a uniform distribution in the interval $\left[ 2 ; 4 \right]$. The sums over the rows are constrained to be equal to $4$ and the sums over the columns are constrained to be equal to $5$. We use raking weights equal to the inverse of the square of the observations. We first apply raking with the $\chi$-square loss. As no other constraint is applied on the raked values, we note in Figure~\ref{fig:distance_effect} that some of the raked values are negative. We then apply raking with the entropic loss. The corresponding raked values are then all positive, as where the initial observations. However, we still have raked values lower than $0.5$. We finally apply the logistic loss with the lower bounds equal to $0.5$ and the upper bounds equal to $4$. The raked values are then all contained in the desired interval.

\section{Strategies for missing data}
\label{sec:example_missing}

To illustrate how missing data can be recovered from the raking as explained in Section 3.3, we take a 3D raking example with cardinality $3 \times 4 \times 5$ (i.e. $60$ estimates to rake), informed by $3$ constraints and $36$ aggregate observations. To obtain a baseline rake estimate (no missingness), we rake the initial $60$ values using constraints and aggregates.  We then remove an entire set of observations for the first dimension, so we have $3$ missing observations and one fewer aggregate that corresponds to these.  

We are interested in two questions. First, how much do the three missing values affect the raking results for the $57$ non-missing values? Second, how well can we recover the three missing observations (compared to the baseline raked results where none are missing)?  

\begin{figure}[h!]
\begin{center}
\includegraphics[width=\textwidth]{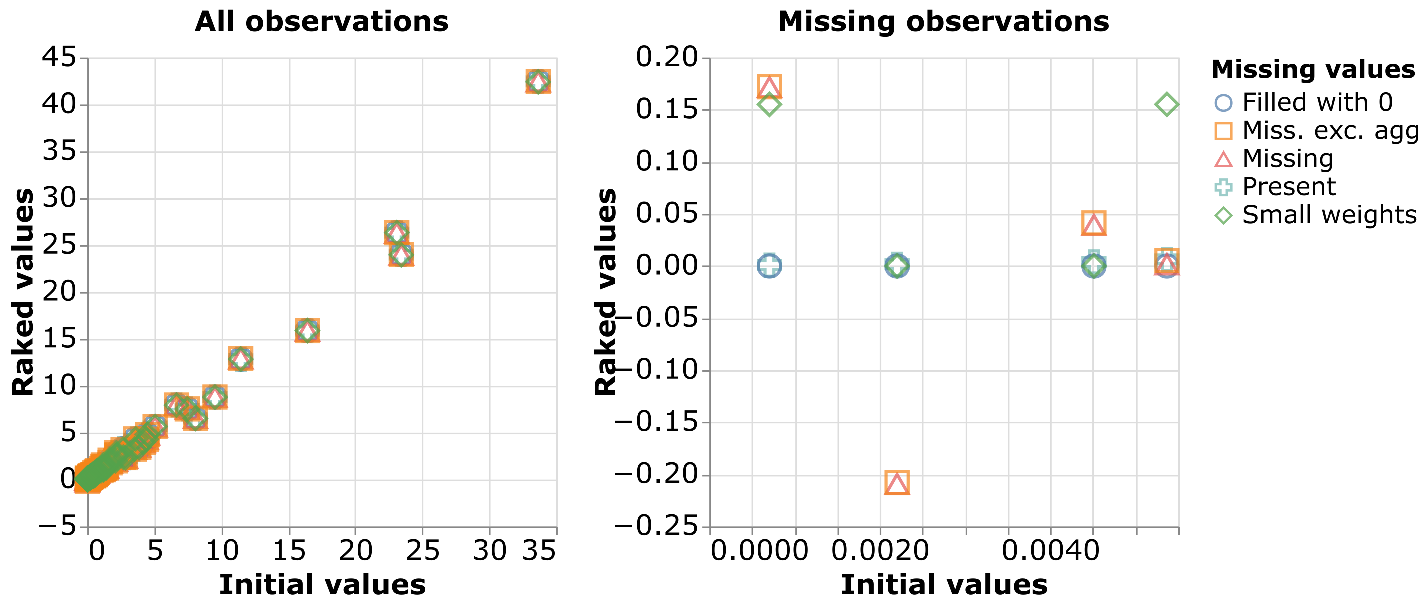}
\end{center}
\caption{Raked versus initial values when filling the missing values with $0$s (blue circles), when computing  the missing observations from the other observations and the constraints and the aggregate is present (orange squares) or not present (red triangles), when giving small weights to the missing values (green diamonds) or when no observations are missing (blue crosses). The raked values corresponding to the known observations are only slightly affected by the treatment of the missing values.}
\label{fig:missing_values}
\end{figure}

We use four scenarios:
\begin{enumerate}
    \item Missing values are replaced by $0$s and treated as observations.
    \item Missing values and their aggregate are treated as missing.
    \item Missing values are treated as missing; their aggregate is available.
    \item Missing values are replaced by average observations value and assigned a small weight.
\end{enumerate}
In Figure~\ref{fig:missing_values}, we compared the raked values obtained 
using these strategies, compared to values we would get when observations are present.

First, raked values corresponding to the known observations are essentially unaffected by the strategy for missing values. 

Second, we used the entropic distance for the raking, which ensures that the observations present in the data set stay positive after the raking. However, if we treat values as truly missing, there is no entropic loss on them, and hence no guarantee that the recovered missing values will also be positive. Imputing a positive value and assigning small weights to the missing values allows us to impose the positivity conditions. 

\begin{frontmatter}
\title{User interface}
\runtitle{User interface}
\end{frontmatter}

Here we describe the simple user interface used to implement different flavors of raking detailed in Section 3. The observations and margins are given by the user through a Pandas data frame containing the following columns:
\begin{itemize}
    \item `value' column:  Value of an observation, a constraint, or `NaN' if missing.
    \item `weight': Positive value for an observation, 0 for missing, or `inf' for constraint.
    \item Categorical values of dimension columns (see below).
\end{itemize}
The user builds the data by providing the list of categorical variables and values corresponding to aggregates over given dimension, and names for value and weight columns. 
For a table $2 \times 2$ with one missing value, the interface thus has the structure given in Table~\ref{tab:interface}.
\begin{table*}[h!]
\caption{Full 2D problem specification. The first four entries specify granular observations, with the NaN value and $0$ weight in the fourth row indicating that the $\left( 2 , 2 \right)$ entry is missing. The last three rows specify aggregates using the special value $0$ (see Data Builder below). Weights of `inf' indicate that the first two aggregates  are set as constraints; while the last row specifies an observation-type aggregate.}
\label{tab:interface}
\begin{tabular}{@{}cccc@{}}
value & X1 & X2 & weights \\
\hline
1.0 & 1 & 1 & 1.0 \\
2.0 & 1 & 2 & 1.0 \\
3.0 & 2 & 1 & 1.0 \\
NaN & 2 & 2 & 0.0 \\
4.0 & 1 & 0 & inf \\
7.0 & 2 & 0 & inf \\
5.0 & 0 & 1 & 10
\end{tabular}
\end{table*}

The user then needs to specify the data builder:
\begin{Verbatim}[frame=single]
data_builder = DataBuilder(
    dim_specs={'X1': 0, 'X2': 0},
    value='value',
    weights='weights')
\end{Verbatim}
For mortality estimates in Section 4, a partial specification is shown in Table~\ref{tab:USHD_interface} and the data builder is defined as below:
\begin{Verbatim}[frame=single]
data_builder = DataBuilder(
    dim_specs={'cause': 0, 'race': 0, 'county': 0},
    value='value',
    weights='weights',
\end{Verbatim}
\begin{table*}[h!]
\caption{Select rows from specification of the Mortality Application in Section 4, showing different levels of aggregates as well as margins and observations.}
\label{tab:USHD_interface}
\begin{tabular}{@{}ccccc@{}}
value & county & race & cause & weights \\
\hline
$y_{001}$ & 1 & 0 & 0 & 1.0 \\
$y_{101}$ & 1 & 0 & 1 & 1.0 \\
$y_{0J1}$ & 1 & J & 0 & 1.0 \\
$y_{IJK}$ & K & J & I & 1.0 \\
$s_{1}$ & 0 & 0 & 1 & inf \\
\end{tabular}
\end{table*}

\end{document}